\documentclass[12pt]{iopart}

\usepackage{iopams}
\usepackage{graphicx}
\usepackage{xcolor}
\usepackage{caption}
\usepackage{subcaption}
\usepackage{hyperref}
\usepackage{etoolbox}
\usepackage{cancel}
\usepackage{cite}

\patchcmd{\numparts}{\addtocounter{equation}{1}}{\refstepcounter{equation}}{}{}

\begin{document}

\title[From ACPs to AMB+]{Field theory of active chiral hard disks: A 
first-principles approach to steric interactions}

\author{Erik Kalz$^1$, Abhinav Sharma$^{2,3}$, Ralf Metzler$^{1,4}$}
\address{$^1$Institute of Physics and Astronomy, University of Potsdam, 
D-14476 Potsdam, Germany}
\address{$^2$Institute of Physics, University of Augsburg, D-86159 Augsburg, 
Germany}
\address{$^3$Institute Theory of Polymers, Leibniz-Institute for Polymer 
Research, D-01069 Dresden, Germany}
\address{$^4$Asia Pacific Centre for Theoretical Physics, KR-37673 Pohang, 
Republic of Korea}
\ead{erik.kalz@uni-potsdam.de}
\ead{abhinav.sharma@uni-a.de}
\ead{ralf.metzler@uni-potsdam.de}

\vspace{10pt}
\begin{indented}
\item[]\today
\end{indented}

\begin{abstract}
A first-principles approach for active chiral hard disks is presented, that 
explicitly accounts for steric interactions on the two-body level. We derive an 
effective one-body equation for the joint probability distribution of positions 
and angles of the particles. By projecting onto the angular modes, we write a 
hierarchy for the lowest hydrodynamic modes, i.e. particle density, polarisation, 
and nematic tensor. Introducing dimensionless variables in the equations, we
highlight the assumptions, which - though inherent - are often included implicit in 
typical closure schemes of the hierarchy. By considering 
different regimes of the P{\'e}clet number, the well-known models in active 
matter can be obtained through our consideration. Explicitly, we derive  
an effective diffusive description and by going to higher 
orders in the closure scheme, we show that this first-principles approach 
results in the recently introduced Active Model B +, a natural extension of the 
Model B for active processes. Remarkably, here we find that chirality can 
change the sign of the phenomenological activity parameters. 
\end{abstract}

\vspace{2pc}

%\submitto{\jpa}

\maketitle

\section{Introduction}

The analytic model of an \textit{Active Chiral Particle (ACP)} 
\textcolor{black}{\cite{van2008dynamics, mijalkov2013sorting, volpe2014simulation,
lowen2016chirality, sevilla2016diffusion}} represents an extension to 
the well-known \textit{Active Brownian Particle (ABP)} model
\cite{schimansky1995structure, schweitzer1998complex, ebeling1999active, 
romanczuk2012active} 
with additional active chirality. These models have proven very 
useful as they come as a first step in generalising equilibrium models to 
include 
non-equilibrium, active contributions for microscopic agents. They constitute a 
minimalist attempt to describe directed motion on the microscale and are found to 
lead to emerging complex structures on the macroscale. For that reason, they are 
widely used in the analytic description of non-equilibrium systems 
\cite{ramaswamy2010mechanics, marchetti2013hydrodynamics, 
julicher2018hydrodynamic}. 

The ABP model describes the overdamped motion of a tagged particle influenced by 
two contributions: \textcolor{black}{firstly, equilibrium fluctuations of the 
surrounding medium, which give rise to the well-known equilibrium Langevin 
description.} 
The particle is assumed to diffuse with a spatial diffusion coefficient $D_T$. 
Secondly, the particle experiences a non-equilibrium driving force, that might 
originate from an internal energy depot or external energy input 
\textcolor{black}{\cite{ebbens2010pursuit, buttinoni2012active, feldmann2019light}}. 
In the simplest case, this 
scenario is modelled as a self-propulsion with a (constant) velocity $v$ along 
an orientation vector $\hat{\mathbf{e}}(\theta)$. In the model of an ABP, the 
self-propulsion direction gets randomised by assuming that the orientation 
vector 
additionally is performing rotational diffusion on the 
unit-sphere with the diffusion coefficient $D_R$. Despite its simplicity, this 
``toy'' ABP model has proven tremendously successful in 
analytical analysis and as a simulation model to describe active matter on a 
macroscale \cite{elgeti2015physics, shaebani2020computational}. Emerging complex 
macroscopic phases such as motility-induced phase separations 
\textcolor{black}{\cite{tailleur2008statistical, fily2012athermal, 
buttinoni2013dynamical, 
palacci2013living, speck2014effective,cates2015motility}},
\textcolor{black}{(chemo-) tactic behaviour \cite{schnitzer1990strategies, 
vuijk2021chemotaxis}},
flocking \cite{levis2019simultaneous, levis2020flocking}, \textcolor{black}{or} 
swarming \cite{vicsek1995novel, czirok2000collective, chate2008modeling} 
have been reported. 

The ABP model has been further developed to include active 
processes that explicitly break the spatial symmetry via an effect of active 
chirality $\omega$ \cite{van2008dynamics, mijalkov2013sorting, 
volpe2014simulation, lowen2016chirality}.
This generalisation in the model allows to describe even more complex 
emerging phenomena, such as odd viscosity \cite{banerjee2017odd, soni2019odd},
finite-size rotating clusters \textcolor{black}{\cite{massana2021arrested, 
ma2022dynamical}}, 
\textcolor{black}{hyperuniform behaviour \cite{lei2019nonequilibrium, 
kuroda2023microscopic}} 
and edge currents at interfaces \textcolor{black}{\cite{liebchen2017collective,
caporusso2024phase, reichhardt2019reversibility}}. This analytical extension is 
inspired by experimental observation of bacteria \cite{berg1990chemotaxis, 
diluzio2005escherichia, lauga2006swimming}, sperm cells \cite{riedel2005self, 
friedrich2007chemotaxis}, or syntactical particles \cite{kummel2013circular, 
arora2021emergent} \textcolor{black}{as well as macroscopic chiral robots 
\cite{scholz2018rotating, yang2020robust, lopez2022chirality}},
 which show an archetypal chirality in their trajectories 
on the microscale. Remarkably, already over a century ago, experimental 
observations were made in living organisms, which showed that trajectories of 
certain microorganisms such as \textit{Loxodes} and  \textit{Paramecium} break 
the spatial symmetry and need an interpretation in terms of chiral 
self-propulsion \cite{jennings1901significance}. 

\textcolor{black}{An alternative, more coarse-grained viewpoint} for describing 
the emerging complex structures in active systems is based on continuum 
field-theoretical models of active matter. The starting point is to address the 
diffusive behaviour of a conserved order parameter, such as the mean particle 
\textit{probability-density function (PDF)} 
\textcolor{black}{$\varrho(\mathbf{x}, t)$, where $\varrho(\mathbf{x}, t)\, 
\mathrm{d}\mathbf{x}$ describes the probability of 
finding a particle in the interval $\mathbf{x}$ and $\mathbf{x} + 
\mathrm{d}\mathbf{x}$ at time $t$}, as in the well-known 
\textit{Model B} \cite{hohenberg1977theory}. This model successfully describes 
the equilibrium behaviour of matter, and especially the dynamics of phase 
separation \cite{bray2002theory}. An inherent assumption in the derivation of 
these field-theoretical models is that the underlying processes follow the 
detailed-balance dynamics, that is broken in active systems. It is, therefore, 
natural that field theories that aim to describe the behaviour of active matter 
have to revisit the foundations and need to go beyond the detailed-balance 
restrictions. Successive works culminated in the recently introduced, so-called 
\textit{Active Model B+ (AMB+)} \textcolor{black}{\cite{tjhung2018cluster, 
cates2023classical,nardini2017entropy}}, that attracted a lot of 
interest lately \textcolor{black}{\cite{speck2022critical, te2023derive,
 zheng2023universal, rapp2019systematic}}.
Continuum approaches, very recently, have also been 
\textcolor{black}{extended} 
to include active chirality and specifically account for broken-time reversal 
symmetries \cite{fruchart2023odd, kuroda2023microscopic, huang2023generalized}. 

The modelling of inter-particle interactions is crucial for such continuum 
approaches. While previous approaches are 
predominantly focused on interaction potentials, here we model the interactions 
via a geometric approach instead. The approach was originally introduced by 
Bruna and Chapman in Refs. \cite{bruna2012diffusion, bruna2012excluded} and 
thereafter successfully applied to ABPs \textcolor{black}{already} 
\cite{bruna2022phase}.
The basic idea is to include particle interactions by restricting the domain of 
definition of the time-evolution equation and thereby creating forbidden areas, 
which correspond to situations with a particle overlap. We apply this idea to 
the ACP model and derive an effective one-body description of the full 
one-particle PDF $p(\mathbf{x}, \theta, t)$ to find a particle at position 
$\mathbf{x}$ with the self-propulsion vector of angle $\theta$ at time $t$ after 
starting with \textcolor{black}{the sharp initial conditions 
$\mathbf{x}=\mathbf{x}_0,\, \theta=\theta_0$, i.e. 
$p(\mathbf{x}, \theta, t_0) = \delta(\mathbf{x}-\mathbf{x}_0)\, 
\delta(\theta - \theta_0)$} 
at $t=t_0$. The resulting time-evolution equation explicitly accounts for 
two-particle steric interactions and therefore its validity is restricted to 
the dilute limit. To arrive at a field-theoretical description for the 
mean particle PDF $\varrho(\mathbf{x}, t)$, one typically proceeds by 
integrating out the effect of the angle \cite{cates2013active, solon2015active}. 
We follow this procedure and arrive at an (infinite) hierarchy of the 
hydrodynamic modes of the active particle, of which the mean particle PDF 
constitutes the zeroth order mode. This hierarchy is mathematically very similar 
to the famed Bogoliubov-Born-Green-Kirkwood-Yvon \textcolor{black}{(BBGKY)} 
hierarchy in kinetic 
theory \cite{cercignani1994mathematical}. Similar to that hierarchy, in 
continuous active matter theories one is also required to close the hierarchy to 
make analytic progress \textcolor{black}{\cite{bertin2006boltzmann, 
bertin2009hydrodynamic}}. 

We transform the closure problem into a perturbation problem and derive two 
field-theoretical descriptions for ACPs, based on the strictness of assumption 
on the perturbation parameters. This gives us access to the otherwise 
phenomenological coefficients in these models. We show that both 
\textcolor{black}{an effective diffusive description} and the AMB+ can be obtained 
within our approach, depending on the order of the closure scheme. 
Our work moreover suggests that the AMB+ is a natural generalisation of 
equilibrium field theories to describe the continuum behaviour of active matter.
We further have first-principles access to the coefficients in the AMB+ and it 
turns out, that, surprisingly, they are altered by active chirality in such a 
way that they can even change sign as a function of chirality. 

The remainder of this work is organised as follows. In Section \ref{model_setup}
we introduce the mathematical model and describe our approach to deal with 
inter-particle interactions in the geometric sense. In Section 
\ref{section_effective_description} we derive in detail the effective one-body 
time-evolution equation of the PDF. To our understanding, the physics community 
is rather unaware of this specific method to handle inter-particle interactions, 
for which we introduce it in appropriate detail. In Section 
\ref{section_hierarchy} 
we thereafter derive the hierarchy of hydrodynamic modes, and in Section 
\ref{section_closure_of_hierarchy} we introduce the mathematical steps to close 
the hierarchy. In Section \ref{section_AMB_plus} we go one step beyond the 
simplest closure and find that the time-evolution of the mean particle PDF 
equals the form predicted by the AMB+. In Section \ref{section_conclusion} 
we conclude and provide an outlook to further and related works. 

\section{Theory: From active chiral particles to the Active Model B +}
\subsection{Model}
\label{model_setup}
In this Section, we introduce the model of interacting ACPs as sketched in 
Fig.~\ref{tikz_graphics_ACP}
and how to deal with their excluded-volume interactions in a geometric sense, 
see also Fig.~\ref{tikz_graphics_excluded_volume}. 
We finish by formulating an effective one-body description. 

\subsubsection{Setup.}
We consider the dynamics of $N$ interacting ACPs in two dimensions. The 
particles centres $\mathbf{x}_i(t)$ and angular coordinates $\theta_i(t)$, 
$i\in\{1, \ldots, N\}$ move according to the overdamped Langevin dynamics 
\cite{sevilla2016diffusion, liebchen2022chiral}
\begin{numparts}
    \label{langevin_description}
\begin{eqnarray}
\label{lanegvin_poistion_setup_acp}
    \frac{\partial}{\partial t}\mathbf{x}_i &= v\, \hat{\mathbf{e}}(\theta_i) + 
    \sqrt{2 D_T}\, \boldsymbol{\eta}_i (t), \\
    \label{model_setup_acp}
    \frac{\partial}{\partial t}\theta_i &= \omega + \sqrt{2 D_R}\, \zeta_i (t).
\end{eqnarray}
\end{numparts}
Here $\boldsymbol{\eta}_i (t)$ and $\zeta_i (t)$ are independent Gaussian white 
noises 
\textcolor{black}{with correlators $\langle \eta_{i, \alpha}(t)
\eta_{j, \beta}(t^\prime)\rangle 
= \delta_{ij}\, \delta_{\alpha\beta}\, \delta(t - t^\prime)$ and 
$\langle \zeta_{i}(t)\zeta_{j}(t^\prime)\rangle 
= \delta_{ij}\, \delta(t - t^\prime)$, where Greek indices $\alpha, \beta$ 
refer to particle coordinates and Latin indices $i,j$ refer to particle labels.} 
$v$ is the (constant) active self-propulsion velocity, $\omega$ the active 
torque and $\hat{\mathbf{e}}(\theta_i) = (\cos(\theta_i), 
\sin(\theta_i))^\mathrm{T}$ is the unit orientation vector, where 
$(\cdot)^\mathrm{T}$ denotes a matrix transpose. Note that for $\omega = 0$, 
the model of Eqs.~\eref{lanegvin_poistion_setup_acp} 
and \eref{model_setup_acp} reduces to the well-known model of ABPs. $D_T$ and 
$D_R$ are the translational and rotational diffusion coefficients, respectively. 
\textcolor{black}{Note that} the diffusion coefficients have different physical 
units, $[D_T] = m^2/s$ and $[D_R] = 1/s$.

The $N$ identical disk-like particles have a diameter $d$ and are assumed to 
interact hard-core with each other. For an illustration of the model system 
see also Fig.~\ref{tikz_graphics_ACP}. The ACPs are modelled to diffuse in a 
spatially bounded domain $\mathbf{x}_i(t) \in \Omega \subset \mathbb{R}^2$ of 
typical size $L \times L$ and their angular coordinates are 
$\theta_i(t) \in [0, 2\pi)$. \textcolor{black}{Using dimensionless quantities 
by rescaling with $L$}, the typical size of the domain is set to unity, whereas 
the diameter of the particle becomes $\varepsilon_d = d/L$. We restrict the 
analysis to a dilute system, i.e., we assume that $N\varepsilon_d \ll 1$. 
In contrast to typical approaches to model the interactions, that is, via an 
interaction-potential term in the spatial Langevin 
equation~\eref{lanegvin_poistion_setup_acp}, 
we instead restrict the domain of definition of the centre-of-mass coordinates 
$\mathbf{x}_i(t)$. We thereby follow a geometric approach to model particle 
interactions as established by Bruna and Chapman \cite{bruna2012diffusion, 
bruna2012excluded} that already has proven successful in various contexts 
\cite{bruna2014diffusion, bruna2015diffusion, kalz2022collisions}. These 
authors also showed in a recent work \cite{bruna2022phase} that within this model 
they can derive additional nonlinear cross-diffusion terms for the description 
of sterically interacting ABPs, compared to more classical treatments of 
interacting ABPs \cite{bialke2013microscopic, speck2015dynamical}. 

The fundamental idea in modelling steric interactions in this geometric 
sense is that Eq.~\eref{lanegvin_poistion_setup_acp} is defined on a restricted 
(spatial) domain $\Omega^N_{\varepsilon_d} = \{ (\mathbf{x}_1, \ldots, 
\mathbf{x}_N) \in \Omega^N; \forall i \neq j\colon |\mathbf{x}_i(t) - 
\mathbf{x}_j(t)| \geq \varepsilon_d\}$ due to the excluded volume. The full 
model including Eq.~\eref{model_setup_acp} therefore is defined on $\Lambda^N 
= \Omega^N_{\varepsilon_d} \times [0, 2\pi)^N$. Note that as the angular 
coordinate is not affected by the steric interactions, its domain of definition 
is not restricted. While for the angular coordinate we assume periodic boundary 
conditions, the cost of treating the particle interactions by restricting the 
domain of definition is that we have generated an inner (moving) boundary as 
apparent from Fig.~\ref{tikz_graphics_ACP}. At this boundary, particles perform 
hard elastic collisions similar to the container walls. Both inner and 
container-wall boundaries therefore are treated with reflective boundaries. To 
simplify this problem, we observe that for a dilute system finding 
configurations in the system where three particles are close, or two particles 
are close to a container wall is of the order $\mathcal{O}(\varepsilon_d^2 N^2)$. 
Configurations where two particles are close or one particle is close to the 
wall, in contrast, are of the order $\mathcal{O}(\varepsilon_d N)$ 
\cite{bruna2012excluded, bruna2012diffusion}. It is therefore reasonable to 
assume that in a dilute system, two-body collisions dominate the interactions 
and we can safely ignore higher-order correlations. As we assume that all 
particles are identical, it is sufficient to consider a system with $N=2$ 
particles. In the end we will scale the resulting interaction contribution by 
the particle number, which is justified as long as we stay within the dilute 
limit. 

\subsubsection{Joint Fokker-Planck equation.}

\begin{figure}
    \centering
\begin{subfigure}{0.45\textwidth}
\centering
    \includegraphics[width=0.9\textwidth]{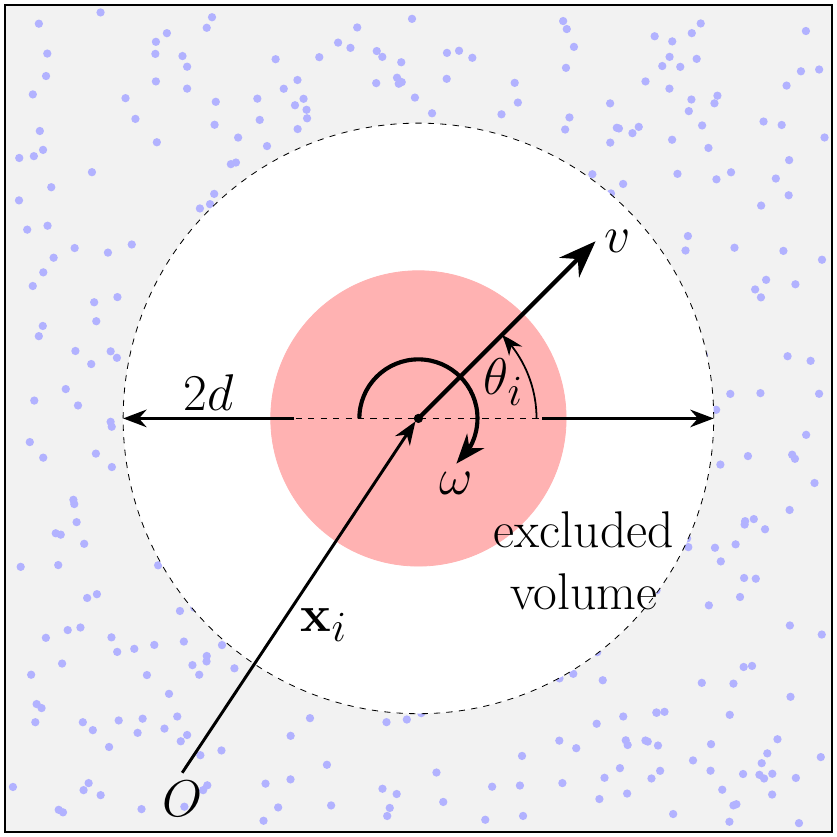}
    \caption{}
    \label{tikz_graphics_ACP}
\end{subfigure}
\begin{subfigure}{0.52\textwidth}
\centering
    \includegraphics[width=0.98\textwidth]{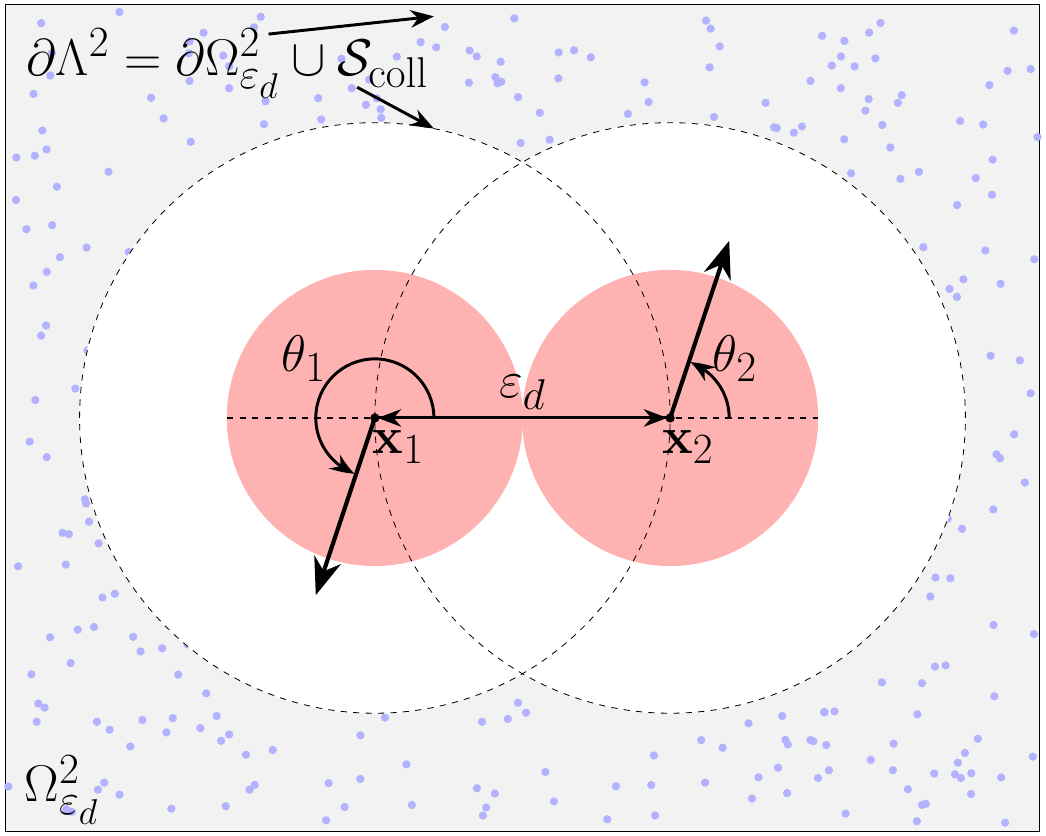}
    \caption{}
    \label{tikz_graphics_excluded_volume}
\end{subfigure}
\caption{(a) Sketch of the model setup of the $i$th spherical ACP of diameter 
$d$. The particle is self-propelled by the active speed $v$ whose 
direction rotates with the active frequency $\omega$. It is embedded in a 
thermal bath giving rise to fluctuations in the spatial and angular coordinates 
$(\mathbf{x}_i, \theta_i)$ with respective strengths $\propto \sqrt{D_T}$ and 
$\propto \sqrt{D_R}$, see Eqs.~\eref{lanegvin_poistion_setup_acp} and 
\eref{model_setup_acp}, respectively. (b) The particle coordinates are defined 
on $\Lambda^2= \Omega_{\varepsilon_d}^2 \times [0,2\pi)^2$, where 
$\Omega_{\varepsilon_d}^2 = \{(\mathbf{x}_1,\mathbf{x}_2) \in \Omega^2; 
|\mathbf{x_1}(t) - \mathbf{x}_2(t)|\geq \varepsilon_d\}$ is the allowed (rescaled)
configuration space \textcolor{black}{within the bounded domain} $\Omega\subset 
\mathbb{R}^2$ \textcolor{black}{and subtracted by} the regions of a particle 
overlap (white area). This excluded-volume creates a reflecting boundary 
\textcolor{black}{on} $\partial \Lambda^2= \partial\Omega^2 \cup 
\mathcal{S}_\mathrm{coll}$, \textcolor{black}{made up from} the container 
walls $\partial\Omega^2$ and an inner moving, so-called collision surface 
$\mathcal{S}_\mathrm{coll}= \{(\mathbf{x}_1,\mathbf{x}_2) \in \Omega^2; 
|\mathbf{x_1}(t) - \mathbf{x}_2(t)| = \varepsilon_d\}$. Note that 
$\varepsilon_d=d/L$, where $L\times L =|\Omega|$ is the typical size of the bounded 
domain $\Omega$.}
\end{figure}

The Fokker-Planck equation (FPE) for the two-body joint PDF $P_2(t) = 
P_2(\mathbf{x}_1, \theta_1, \mathbf{x}_2, \theta_2, t)$, defined on $\Lambda^2$, 
reads \cite{risken1996fokker}
\begin{numparts}
\begin{eqnarray}
\label{ACP_N_particle_FP_equation}
    \frac{\partial }{\partial t} P_2(t) &= \nabla_1 \cdot \bigg[D_T 
    \nabla_1 - v\hat{\mathbf{e}}(\theta_1) \bigg] P_2(t) +  
    \frac{\partial}{\partial \theta_1} \left[D_R 
    \frac{\partial}{\partial \theta_1} - \omega \right]P_2(t) \nonumber\\ 
    &\quad + \nabla_2 \cdot \bigg[D_T \nabla_2 - v\hat{\mathbf{e}}(\theta_2) 
    \bigg] P_2(t) +  \frac{\partial}{\partial \theta_2} \left[D_R 
    \frac{\partial}{\partial \theta_2} - \omega \right]P_2(t).
\end{eqnarray}
Here $\nabla_i$ denotes the partial differential vector operator with respect 
to the position of particle $i \in\{1,2\}$. The reflective boundary condition 
reads
\begin{equation}
\label{spatial_bc}
    \mathbf{n}_1 \cdot \bigg[D_T \nabla_1 - v\hat{\mathbf{e}}(\theta_1) \bigg] 
    P_2(t) + \mathbf{n}_2 \cdot \bigg[D_T \nabla_2 - v\hat{\mathbf{e}}(\theta_2) 
    \bigg] P_2(t) = 0, 
\end{equation}
valid on $\partial \Lambda^2 = \partial\Omega_{\varepsilon_d}^2 = 
\partial\Omega^2 \cup \mathcal{S}_\mathrm{coll}$, where 
$\mathcal{S}_\mathrm{coll} = \{ (\mathbf{x}_1,\mathbf{x}_2) \in \Omega^2; 
|\mathbf{x}_1(t) - \mathbf{x}_2(t)| = \varepsilon_d\}$ is the so-called (inner) 
\textit{collision surface}. For an illustration see 
Fig.~\ref{tikz_graphics_excluded_volume}. $\mathbf{n}_i$ in 
Eq.~\eref{spatial_bc} is the outward unit normal vector of disk $i$. Note 
that $\mathbf{n}_i=\mathbf{0}$ for $\mathbf{x}_j \in \partial\Omega^2$ for 
$(i,j) = (1,2)$ and $(2,1)$ due to particle conservation, as well as 
$\mathbf{n}_1= - \mathbf{n}_2$ on $\mathcal{S}_\mathrm{coll}$, due to elastic 
collisions of particles. For the angular coordinate 
Eq.~\eref{ACP_N_particle_FP_equation} is supplemented with the periodic 
boundary condition
\begin{equation}
\label{angular_bc}
    P_{\textcolor{black}{2}}(\theta_i = 0) = P_{\textcolor{black}{2}}
    (\theta_i = 2\pi).
\end{equation}
\end{numparts}

It is convenient to use the structural similarities of the spatial and angular 
coordinates and write Eq.~\eref{ACP_N_particle_FP_equation} in terms of joint 
variables $\chi_i= (\mathbf{x}_i, \theta_i)$. The diffusion coefficients form a 
diffusion matrix $\mathbf{D} = \mathrm{diag}(D_T, D_T, D_R)$ and the joint drift 
reads $\mathbf{f}(\theta_i) = (v\ \hat{\mathbf{e}}(\theta_i), 
\omega)^\mathrm{T}$. The FPE  \eref{ACP_N_particle_FP_equation} written in the 
joint variables becomes
\begin{eqnarray}
\label{ACP_N_particle_FP_equation_joint}
    \frac{\partial }{\partial t} P_2(t) = \nabla_{\chi_1} \cdot 
    \left[\mathbf{D}\nabla_{\chi_1} - \mathbf{f}(\theta_1) \right] P_2(t) + 
    \nabla_{\chi_2} \cdot \left[\mathbf{D}\nabla_{\chi_2} - 
    \mathbf{f}(\theta_2) \right]  P_2(t),
\end{eqnarray}
valid on $\Lambda^2$. We here use the joint partial differential operator 
$\nabla_{\chi_i} = (\nabla_i, \partial / \partial \theta_i)^\mathrm{T}$.

We are interested in analytically capturing the effects of particle collisions 
on the one-body diffusive behaviour. Therefore, we aim at deriving an effective 
description for the full one-body PDF $p(\mathbf{x}_1, \theta_1, t) = 
p(\chi_1, t)$, which is defined as
\begin{eqnarray}
    p(\chi_1, t) = \int_{\Lambda(\chi_1)}\mathrm{d}\chi_2\ P_2(t) = 
    \int_0^{2\pi}\mathrm{d}\theta_2 \int_{\Omega \setminus 
    \mathrm{B}_{\varepsilon_d}(\mathbf{x}_1)}\mathrm{d}\mathbf{x}_2\ P_2(t), 
\end{eqnarray}
where we take particle one to be the test particle of interest. The area 
$\Lambda(\chi_1)$ of integration is given by all allowed configurations for 
the second particle to be placed everywhere apart from the excluded volume 
created by the first particle, i.e. $\Lambda(\chi_1) = \Omega \setminus 
\mathrm{B}_{\varepsilon_d}(\mathbf{x}_1) \times [0,2\pi)$, where 
$\mathrm{B}_{\varepsilon_d}(\mathbf{x}_1)$ is the disk of radius 
$\varepsilon_d$ centred at $\mathbf{x}_1$.

\subsection{Effective one-body description}
\label{section_effective_description}

In this Section, we perform the integration to arrive at an effective one-body 
description and encounter that the effect of particle collisions results in a 
so-called collision integral. We solve this integral by the method of matched 
asymptotic expansions and provide the effective one-body description. This 
Section follows Ref.~\cite{bruna2022phase} and adapts it to ACPs to review and 
introduce the geometric method to deal with particle interactions to the reader.

\subsubsection{The collision integral.}

Integrating Eq.~\eref{ACP_N_particle_FP_equation} over the reduced 
configuration space $\Lambda(\chi_1)$ results in
\begin{eqnarray}
\label{starting_integral}
    \frac{\partial}{\partial t} p(\chi_1, t) &= 
    \int_{\Lambda(\chi_1)}\mathrm{d}\chi_2\ \nabla_{\chi_1} \cdot 
    \bigg[\mathbf{D} \nabla_{\chi_1} - \mathbf{f}(\theta_1) \bigg] P_2(t) 
    \nonumber \\  
     &\quad + \int_{\Lambda(\chi_1)}\mathrm{d}\chi_2\ \nabla_{\chi_2} \cdot 
     \bigg[\mathbf{D} \nabla_{\chi_2} - \mathbf{f}(\theta_2)\bigg] P_2(t).
\end{eqnarray}
We can easily evaluate the second integral, in which integration and 
differentiation are with respect to the same variable. Using the divergence 
theorem and applying the boundary conditions of Eqs.~\eref{spatial_bc} and 
\eref{angular_bc}, we find
\begin{eqnarray}
    \label{applied_normal_vector}
    \int_{\Lambda(\chi_1)}\mathrm{d}\chi_2\ \nabla_{\chi_2} \cdot 
    \bigg[\mathbf{D} \nabla_{\chi_2} - \mathbf{f}(\theta_2) \bigg] P_2(t) 
    \nonumber\\ = \int_0^{2\pi}\mathrm{d}\theta_2\int_{\partial
    \mathrm{B}_{\varepsilon_d}(\mathbf{x}_1)}\mathrm{d}\mathrm{S}_{2}\ 
    \mathbf{n}_2 \cdot \bigg[D_T \nabla_{1} - v \hat{\mathbf{e}}(\theta_1) 
    \bigg]P_2(t), 
\end{eqnarray}
where $\mathrm{d}\mathrm{S}_{2}\, \mathbf{n}_2$ is the outward surface element 
of $\mathbf{x}_2$ and we used that $\mathbf{n}_1 = - \mathbf{n}_2$ on 
$\partial\mathrm{B}_{\varepsilon_d}(\mathbf{x}_1)$ and $\mathbf{n}_1 = 
\mathbf{0}$ for  $\mathbf{x}_2 \in \partial\Omega$.

The first integral in Eq.~\eref{starting_integral} cannot be evaluated that 
simply since integration and differentiation are with respect to different 
particle labels. Instead, we have to use the Reynolds transport theorem 
extended to spatial variation of integrals \cite{kalz2022diffusion, 
bruna_thesis}. Together with an additional use of the divergence theorem 
this results in 
\begin{eqnarray}
\label{seperated_integral_appendix}
\int_{\Lambda(\chi_1)}\mathrm{d}\chi_2\ \nabla_{\chi_1} \cdot \bigg[\mathbf{D} 
\nabla_{\chi_1} - \mathbf{f}(\theta_1) \bigg] P_2(t) =\nabla_{\chi_1} \cdot 
\left[\mathbf{D} \nabla_{\chi_1} - \mathbf{f}(\theta_1) \right] p(\chi_1, t) 
\nonumber \\
\quad - \int_0^{2\pi}\mathrm{d}\theta_2\int_{\partial\mathrm{B}_{\varepsilon_d}
(\mathbf{x}_1)}\mathrm{d}\mathrm{S}_{2}\ \mathbf{n}_2 \cdot \bigg[D_T 
\left(2\nabla_{1} + \nabla_2\right) + v \hat{\mathbf{e}}(\theta_1) \bigg]  
P_2(t).
\end{eqnarray}
For details of this calculation, see \ref{appendix_integral_evaluation}. 
We combine this integral with Eq.~\eref{applied_normal_vector} and find the 
effective equation for the full one-body PDF $p(\chi_1, t)$ according to 
Eq.~\eref{starting_integral} as
\begin{eqnarray}
\label{first_order_equation_with_collision_integral}
    \frac{\partial}{\partial t} p(\chi_1, t) &= \nabla_{\chi_1} \cdot 
    \left[\mathbf{D} \nabla_{\chi_1}  - \mathbf{f}(\theta_1) \right] 
    p(\chi_1, t) \nonumber\\ 
    &\quad -D_T \int_0^{2\pi}\mathrm{d}\theta_2 \int_{\partial
    \mathrm{B}_{\varepsilon_d}(\mathbf{x}_1)}\mathrm{d}\mathrm{S}_{2}\ 
    \mathbf{n}_2 \cdot \left(\nabla_{1} + \nabla_{2} \right)P_2(\chi_1, 
    \chi_2, t).
\end{eqnarray}

In analogy to kinetic theory \cite{cercignani1994mathematical}, we refer to 
this integral as the \textit{collision integral} and denote it by 
$I(\chi_1, t)$. It captures the effect of two-body hard-disk collisions on a 
probabilistic level. To evaluate this integral, we have to find an expression 
for the joint PDF $P_2(\chi_1, \chi_2, t)$ in terms of the one-body PDF 
$p(\chi_1, t)$, similar to the classical closure problem in kinetic theory, 
known as the BBGKY hierarchy. This relation will be specifically relevant in 
regions where the particles are close, and where evaluating the collision 
integral will contribute to the effective description.

\subsubsection{Matched asymptotic expansion.}

We aim at an approximation for $P_2$ via the method of matched asymptotic 
expansion \cite{bruna2022phase, bruna2012diffusion, bruna2012excluded, 
bruna2014diffusion,  bender1999advanced}. We can suppose that when the 
particles are far apart they are independent, given the hard-disk nature of 
the interaction. In this so-called \textit{outer region}, we define the 
\textit{outer joint PDF} as $P^\mathrm{out}(\chi_1,\chi_2,t) = 
P_2(\chi_1,\chi_2,t)$ and due to the independency argument we find that
\begin{equation}
    P^\mathrm{out}(\chi_1,\chi_2,t) = p(\chi_1,t)\ p(\chi_2,t) + \varepsilon_d 
    P^\mathrm{out}_{(1)}(\chi_1,\chi_2,t) + \mathcal{O}(\varepsilon_d^2).
\end{equation}
Note that $P^\mathrm{out}_{(1)}$ is a function denoting the corrections at 
first order $\mathcal{O}(\varepsilon_d)$ to the independence argument 
\cite{bruna2017diffusion}.

\begin{figure}
    \centering
\begin{subfigure}{0.475\textwidth}
\centering
    \includegraphics[width=\textwidth]{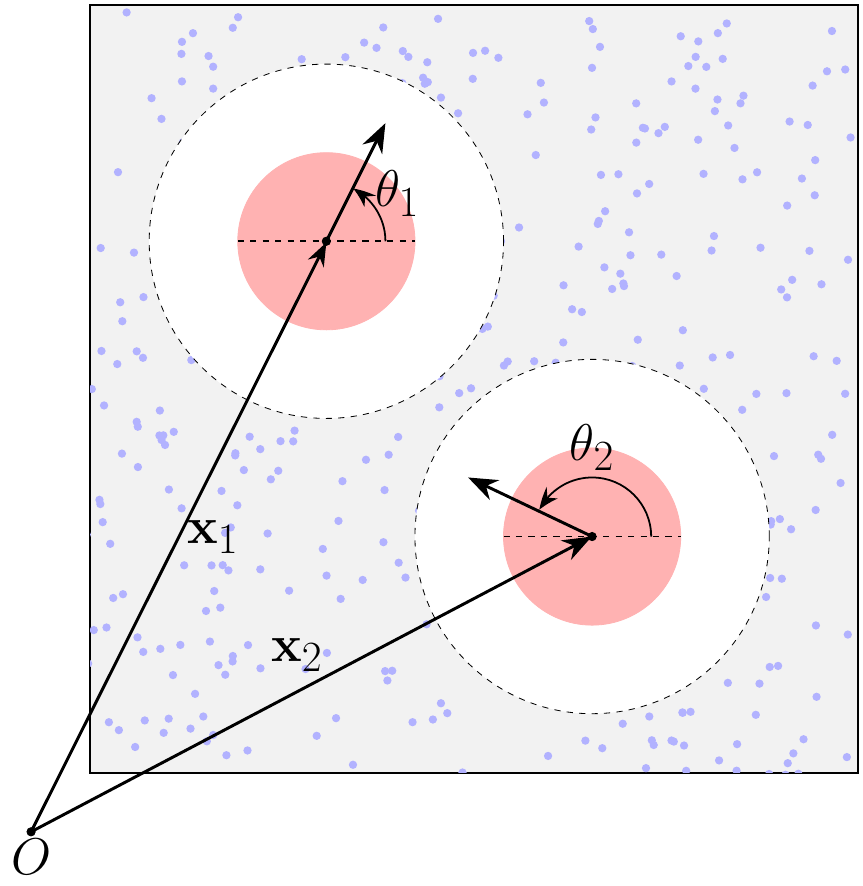}
    \caption{}
\end{subfigure}
\begin{subfigure}{0.475\textwidth}
\centering
    \includegraphics[width=\textwidth]{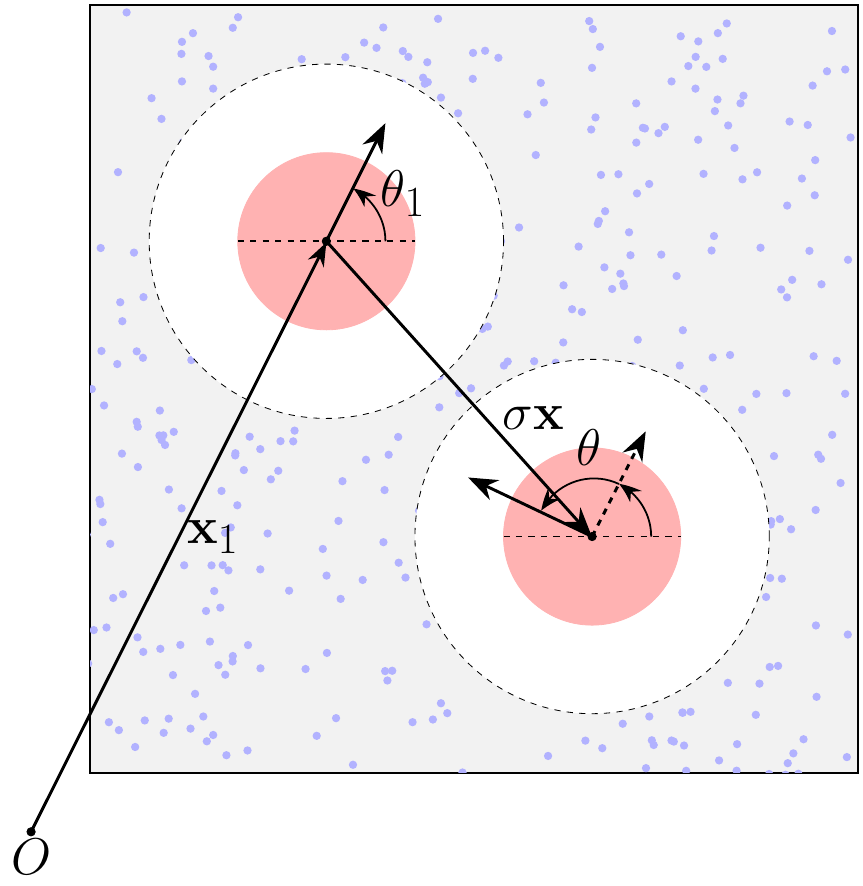}
    \caption{}
\end{subfigure}
\caption{The description of the problem changes from a particle-focused to a 
fixed-particle perspective. The second particle is measured in relative units. 
Technically we change from a description of $(\mathbf{x}_1, \theta_1)$ 
for particle one and $(\mathbf{x}_2, \theta_2)$ for particle two in (a) to a 
description of $(\mathbf{x}_1, \theta_1)$ for the effectively fixed 
particle one and the relative coordinate $(\mathbf{x}_1 + 
\varepsilon_d\mathbf{x}, \theta_1 + \theta)$ for particle two in (b). One advantage 
of this coordinate change is that the collision surface in the new coordinates 
is given by the condition $|\mathbf{x}| = 1$.}
\label{tikz_graphics_coordinate_change}
\end{figure}

In contrast, when the two particles are close, the particles are correlated 
due to interactions and we perform a variable change in this so-called 
\textit{inner region}. We fix particle one and measure the distance to this 
particle with respect to $\varepsilon_d$, see 
Fig.~\ref{tikz_graphics_coordinate_change}. The coordinate change reads as 
$(\textcolor{black}{\chi_1}, \mathbf{x}_2, \theta_2) \mapsto (\chi_1, 
\mathbf{x}_1 + \varepsilon_d\mathbf{x}, \theta_1 + \theta)$.  The 
\textit{inner PDF} is defined as $P^\mathrm{in}(\chi_1, \mathbf{x}, \theta, t) 
= P_2(\chi_1, \chi_2, t)$. Rewriting the two-particle problem into inner 
coordinates, we find
\begin{numparts}
\begin{eqnarray}
\label{full_inner_problem}
    \varepsilon_d^2 \frac{\partial}{\partial t} P^\mathrm{in} &= 2 D_T\ 
    \nabla_{\mathbf{x}}^2 P^\mathrm{in} +\varepsilon_d \nabla_{\mathbf{x}} 
    \cdot \bigg[ v \left(\hat{\mathbf{e}}(\theta_1) - \hat{\mathbf{e}}
    (\theta_1 + \theta)\right) - 2D_T\ \nabla_{\mathbf{x}_1} \bigg] 
    P^\mathrm{in} \nonumber\\ 
    &\quad+ \varepsilon_d^2 \nabla_{\mathbf{x}} \cdot \left[D_T\ 
    \nabla_{\mathbf{x}}  - v\ \hat{\mathbf{e}}(\theta_1) \right] P^\mathrm{in} 
    \nonumber\\ 
    &\quad+ \varepsilon_d^2 \left[D_R\left(\frac{\partial}{\partial \theta_1} - 
    \frac{\partial}{\partial \theta}\right)^2 - \omega 
    \frac{\partial}{\partial \theta_1} \right] P^\mathrm{in}.
\end{eqnarray}
The no-flux boundary condition then translates into inner coordinates as 
\begin{equation}
\label{inner_no_flux_bc}
    2D_T\ \mathbf{x}\cdot \nabla_{\mathbf{x}} P^\mathrm{in} = 
    \varepsilon_d\mathbf{x}\cdot \left[D_T\ \nabla_{\mathbf{x}_1} - 
    v \left(\hat{\mathbf{e}}(\theta_1) - \hat{\mathbf{e}}(\theta_1 + 
    \theta)\right) \right]P^\mathrm{in},
\end{equation}
which is valid on the collision surface $\mathcal{S}_\mathrm{coll}$ defined in 
inner coordinates by $\{\mathbf{x}\in\mathbb{R}^2; |\mathbf{x}|=1\}$. When 
formulating the problem in inner coordinates, we have an additional boundary 
condition, which is replacing the otherwise implicit natural boundary condition 
on $P_2$, i.e., $\lim_{|\mathbf{x}_i| \to \infty} P_2 = 0, i \in\{1,2\}$. 
In the framework of the matched asymptotic expansion, this condition implies 
that the inner PDF has to match the outer PDF as $|\mathbf{x}| \to \infty$. 
An expansion of the outer solution in inner coordinates gives
\begin{equation}
\label{outer_bc}
    P^\mathrm{out}(\chi_1, \chi_2, t) \textcolor{black}{=} p\ p^+ + \varepsilon_d
    \left(p\ \mathbf{x} \cdot \nabla_{\mathbf{x}_1} p^+ + 
    P^{\mathrm{out},+}_{(1)}\right) \textcolor{black}{+\mathcal{O}
    (\varepsilon_d^2)}, 
\end{equation}
\end{numparts}
where we have introduced the shorthand notations $p = p(\chi_1, t),\ p^+ = 
p(\mathbf{x}_1, \theta_1 + \theta, t)$ and $P^{\mathrm{out},+}_{(1)} = 
P^\mathrm{out}_{(1)}(\chi_1, \mathbf{x}_1, \theta_1 + \theta, t)$ 
\cite{bruna2022phase}. 
Expanding the inner solution in powers of $\varepsilon_d$, $P^\mathrm{in} 
\textcolor{black}{=} P^\mathrm{in}_{(0)} + \varepsilon_d P^\mathrm{in}_{(1)} + 
\mathcal{O}(\varepsilon_d^2)$, the zeroth-order inner problem becomes 
\begin{numparts}
\begin{eqnarray}
    0 = 2D_T\ \nabla_{\mathbf{x}}^2 P^\mathrm{in}_{(0)}, & \\
    0 =  2D_T\ \mathbf{x} \cdot \nabla_{\mathbf{x}} P^\mathrm{in}_{(0)},  
    & \qquad \mathrm{on}\ |\mathbf{x}| = 1, \\ 
    P^\mathrm{in}_{(0)} \sim p p^+, & \qquad\mathrm{as}\ |\mathbf{x}| 
    \to \infty,
\end{eqnarray}
\end{numparts}
with the straight-forward solution 
\begin{equation}
P^\mathrm{in}_{(0)} =  p\ p^+.
\end{equation}
Note here, that $P^\mathrm{in}_{(0)}$ is no function of the separation 
coordinate $\mathbf{x}$. 

Before we can write the first-order inner problem, we note from 
Eq.~\eref{full_inner_problem}, that at order $\mathcal{O}(\varepsilon_d)$,  
we have that
\begin{eqnarray}
    0 &= 2 D_T\ \nabla_{\mathbf{x}}^2  P^\mathrm{in}_{(1)} +\varepsilon_d 
    \nabla_{\mathbf{x}} \cdot \bigg[ v \left(\hat{\mathbf{e}}(\theta_1) - 
    \hat{\mathbf{e}}(\theta_1 + \theta)\right) - 2D_T\ \nabla_{\mathbf{x}_1} 
    \bigg] P^\mathrm{in}_{(0)} \nonumber\\ 
    &= 2D_T\ \nabla_{\mathbf{x}}^2  P^\mathrm{in}_{(1)},
\end{eqnarray}
as $P^\mathrm{in}_{(0)}$ is no function of the separation variable $\mathbf{x}$.
Taking this into account, the first-order inner problem reads
\begin{numparts}
\label{first_order_inner_problem}
\begin{eqnarray}
    0 = 2D_T\ \nabla_{\mathbf{x}}^2  P^\mathrm{in}_{(1)},& \\
    \mathbf{x}\cdot \nabla_{\mathbf{x}} P^\mathrm{in}_{(1)} =  \mathbf{x}\cdot 
    \mathbf{A}(\chi_1, \theta,t), & \qquad \mathrm{on}\ |\mathbf{x}| = 1, \\ 
    \label{first_order_inner_problem_bc_infty}
    P^\mathrm{in}_{(1)} \sim \mathbf{x}\cdot \mathbf{B}(\chi_1, \theta,t) + 
    P^{\mathrm{out},+}_{(1)}, & \qquad \mathrm{as}\ |\mathbf{x}| \to \infty,
\end{eqnarray}
\end{numparts}
where
\begin{eqnarray}
    \mathbf{A}(\chi_1, \theta,t) &= \frac{1}{2D_T} \Big(D_T\ 
    \nabla_{\mathbf{x}_1} (p p^+) - v (\hat{\mathbf{e}}(\theta_1) - 
    \hat{\mathbf{e}}(\theta_1 + \theta)) p p^+\Big), \\
    \mathbf{B}(\chi_1, \theta,t) &= p\ \nabla_{\mathbf{x}_1} p^+.
\end{eqnarray}
The solution can be obtained straightforwardly and is given by 
\cite{bruna2012diffusion, bruna2012excluded, kalz2022collisions, 
kalz2022diffusion}
\begin{equation}
    P^\mathrm{in}_{(1)} = a + P^{\mathrm{out},+}_{(1)} + \mathbf{x}\cdot 
    \mathbf{B} - \frac{\mathbf{x}}{|\mathbf{x}|^2}\cdot \left(\mathbf{A} - 
    \mathbf{B}\right),
\end{equation}
where $a$ is an arbitrary integration constant. According to the expansion 
ansatz $P^\mathrm{in} \sim P^\mathrm{in}_{(0)} + \varepsilon_d 
P^\mathrm{in}_{(1)} + \mathcal{O}(\varepsilon_d^2)$ for the inner PDF, we thus 
found that 
\begin{eqnarray}
\label{final_inner_solution}
P^\mathrm{in} &= p p^+ + \varepsilon_d \mathbf{x} \cdot p\ \nabla_{\mathbf{x}_1} 
p^+ - \varepsilon_d \frac{\mathbf{x}}{2|\mathbf{x}|^2} \cdot \Big[p^+\ 
\nabla_{\mathbf{x}_1} p - p\ \nabla_{\mathbf{x}_1} p^+ \nonumber\\
&\quad- \frac{v}{D_T}\left(\hat{\mathbf{e}}(\theta_1) - \hat{\mathbf{e}}
(\theta_1 + \theta)\right)p p^+ \Big] + \varepsilon_d \big(a + 
P^{\mathrm{out},+}_{(1)}\big) + \mathcal{O}(\varepsilon_d^2).
\end{eqnarray}

\subsubsection{Evaluation of the collision integral.}

We now use this approximate inner solution to evaluate the collision integral. 
In inner coordinates, this is given by 
\begin{equation}
    I(\chi_1, t) = \varepsilon_d  D_T \int_0^{2\pi}\mathrm{d}\theta 
    \int_{|\mathbf{x}| = 1}\mathrm{d}\mathrm{S}_{\mathbf{x}}\ \mathbf{x} 
    \cdot \nabla_{\mathbf{x}_1} P^\mathrm{in}(\chi_1, \mathbf{x}, \theta, t),
\end{equation}
where we used that $\mathbf{n}_{\mathbf{x}} = - \mathbf{x}$ on the collision 
surface $\mathcal{S}_\mathrm{coll}$. Using the inner solution of 
Eq.~\eref{final_inner_solution}, the collision integral becomes
\begin{eqnarray}
\label{unfinished_collision_integral}
    I(\chi_1, t) &= \varepsilon_d D_T \int_0^{2\pi}\mathrm{d}\theta 
    \left[\nabla_{\mathbf{x}_1} (p p^+) \right]_\alpha\ 
    \int_{|\mathbf{x}| = 1} \mathrm{d}\mathrm{S}_{\mathbf{x}}\ x_\alpha \\
    &\quad+ \varepsilon_d^2 D_T \int_0^{2\pi}\mathrm{d}\theta 
    \bigg[\nabla_{\mathbf{x}_1} P^{\mathrm{out},+}_{(1)} \bigg]_\alpha\ 
    \int_{|\mathbf{x}| = 1} \mathrm{d}\mathrm{S}_{\mathbf{x}}\ x_\alpha 
    \nonumber\\
    &\quad+ \varepsilon_d^2 D_T \int_0^{2\pi}\mathrm{d}\theta 
    \bigg[\nabla_{\mathbf{x}_1} (p \nabla_{\mathbf{x}_1} p^+)_\alpha 
    \bigg]_\beta\ \int_{|\mathbf{x}| = 1} \mathrm{d}\mathrm{S}_{\mathbf{x}}\ 
    x_\alpha x_\beta  \nonumber\\
    &\quad- \varepsilon_d^2\frac{D_T}{2} \int_0^{2\pi}\mathrm{d}\theta 
    \bigg[\nabla_{\mathbf{x}_1} (p^+\ \nabla_{\mathbf{x}_1} p - p\ 
    \nabla_{\mathbf{x}_1} p^+ )_\alpha \bigg]_\beta\ \int_{|\mathbf{x}| = 1} 
    \mathrm{d}\mathrm{S}_{\mathbf{x}}\ \frac{x_\alpha x_\beta}{|\mathbf{x}|^2} 
    \nonumber\\
    &\quad- \varepsilon_d^2\frac{v}{2} \int_0^{2\pi}\mathrm{d}\theta 
    \bigg[\hat{\mathbf{e}}(\theta_1) - \hat{\mathbf{e}}(\theta_1 + \theta)
    \bigg]_\alpha \bigg[\nabla_{\mathbf{x}_1} (p p^+)\bigg]_\beta\ 
    \int_{|\mathbf{x}| = 1} \mathrm{d}\mathrm{S}_{\mathbf{x}}\ 
    \frac{x_\alpha x_\beta}{|\mathbf{x}|^2} \nonumber,
\end{eqnarray}
where we introduced the Einstein convention, i.e., that the sum over double 
indices is implicit: $x_\alpha x_\alpha = \sum_{\alpha=1}^2 x_\alpha x_\alpha$. 
There are two types of integrals appearing in 
Eq.~\eref{unfinished_collision_integral}, namely, an integral of an outer unit 
normal vector over the whole unit sphere: $ \int_{|\mathbf{x}| = 1} 
\mathrm{d}\mathrm{S}_{\mathbf{x}}\ x_\alpha= 0$, by geometrical insight; and 
\begin{equation}
    \int_{|\mathbf{x}| = 1} \mathrm{d}\mathrm{S}_{\mathbf{x}}\ 
    \frac{x_\alpha x_\beta}{|\mathbf{x}|^2} = \int_{|\mathbf{x}| = 1} 
    \mathrm{d}\mathrm{S}_{\mathbf{x}}\ x_\alpha x_\beta = \pi\ 
    \delta_{\alpha\beta},
\end{equation}
valid in two dimensions. 

As $\theta_1$ is independent of $\theta$ and kept constant for the 
variation of $\theta$, we can define a new variable $\tilde{\theta} = 
\theta_1 + \theta$ and use it to integrate out the $\theta$-dependence in 
Eq.~\eref{unfinished_collision_integral}. Therefore, we define the 
\textit{mean particle PDF}
\begin{equation}
\label{ad_hoc_definition_mean_density}
    \varrho(\mathbf{x}, t) = \int_0^{2\pi}\mathrm{d}\tilde{\theta}\ 
    p(\mathbf{x}, \tilde{\theta}, t),
\end{equation}
and the \textit{polarisation}
\begin{equation}
\label{ad_hoc_definition_polarisation}
    \boldsymbol{\sigma}(\mathbf{x}, t) = 2 \int_0^{2\pi}\mathrm{d}
    \tilde{\theta}\ p(\mathbf{x}, \tilde{\theta}, t)\ \hat{\mathbf{e}}
    (\tilde{\theta}),
\end{equation}
as the zeroth and first-order moment of the full one-body PDF, respectively. 
Note the factor of $2$ in the definition of the polarisation, which is 
necessary for consistency later on. Thus, the collision integral becomes
\begin{equation}
    I(\chi_1, t) = \varepsilon_d^2\frac{\pi}{2}\ \nabla_{\mathbf{x}_1} \cdot 
    \Big[3D_T p\ \nabla_{\mathbf{x}_1} \varrho - D_T \varrho\ 
    \nabla_{\mathbf{x}_1} p + v\left(\hat{\mathbf{e}}(\theta_1) \varrho - 
    \frac{\boldsymbol{\sigma}}{2} \right)p \Big], 
\end{equation}
where $\varrho = \varrho (\mathbf{x}_1, t),\ \boldsymbol{\sigma} = 
\boldsymbol{\sigma} (\mathbf{x}_1, t)$, and, as a reminder, $p = 
p(\mathbf{x}_1, \theta_1, t)$. 

As expected, the dependence on the separation coordinates $\mathbf{x}$ and 
$\theta$ vanished, since we integrated out the effect of the second particle 
on the first. Thus, we can back-transform into the original variables and then 
drop the index in the notation $(\mathbf{x}_1, \theta_1) \mapsto (\mathbf{x}, 
\theta)$, similarly for the operator $\nabla_{1} \mapsto \nabla$. We insert 
the evaluated collision integral into 
Eq.~\eref{first_order_equation_with_collision_integral} and find the 
effective time-evolution equation for the full one-body PDF $p$. This 
equation is valid for two hard-interacting particles. As introduced in the 
beginning, in the dilute system of $N$ particles, we can safely assume that 
two-particle collisions dominate any higher-order correlations. The tagged 
particle can have $(N-1)$ inner regions with each of the remaining particles 
under this assumption. It is thus sufficient to multiply the collision integral 
by the factor of $(N-1)$ to account for the interaction effect in the 
effective, dilute one-body description. 

\subsubsection{Effective one-body equation.}

We introduce the dimensionless parameter $\phi = 
\textcolor{black}{\varepsilon_d^2} \pi(N-1)/4$, which for large $N$ 
approximately equals the area fraction of the particles.  We then write the 
obtained time-evolution equation for the full one-body PDF $p = p(\mathbf{x}, 
\theta, t)$ as
\begin{eqnarray}
\label{final_full_one_body_equation}
     \frac{\partial}{\partial t} p  &= - v\ \nabla \cdot \bigg[
        ( 1 - 2\phi\varrho)\ \hat{\mathbf{e}} + \phi \boldsymbol{\sigma} 
        \bigg]p + \frac{\partial}{\partial \theta} \left[D_R 
        \frac{\partial}{\partial \theta} - \omega\right] p \nonumber\\ 
     &\quad +D_T\ \nabla \cdot \bigg[(1 - 2\phi \varrho)\ \nabla p  + 6\phi p\ 
     \nabla \varrho \bigg], 
\end{eqnarray}
valid in the dilute limit, i.e., $\phi \ll 1$. Without the effect of chirality 
($\omega =0$) this equation was recently derived in Ref.~\cite{bruna2022phase} 
for ABPs. This systematic derivation of the one-body description results in 
additional cross-diffusion terms ($\propto \varrho\nabla p$, $\propto 
p\nabla\varrho$). \textcolor{black}{These terms were not reported in literature 
on phenomenological approaches \cite{bialke2013microscopic, speck2015dynamical},
works relying on approximations of the pair correlation 
function \cite{bickmann2020predictive,te2023derive}, or classical dynamical 
density functional theory \cite{te2020classical, archer2004dynamical}. These 
cross-diffusion, however, become specifically important when dealing with 
different particle labels, and thus accounting for particle-identity in the 
effective description \cite{bruna2012diffusion,kalz2022collisions}}. Further, 
recent work shows that for ABPs, Eq.~\eref{final_full_one_body_equation} 
forms a well-posed problem for which a stationary state exists 
\cite{burger2023wellposedness}. 

In the context of phase-separating active matter, typically effective 
self-propulsion velocities $v(\varrho)$ are introduced 
\cite{tailleur2008statistical, stenhammar2013continuum} and constitute an 
essential theoretical trick to achieve phase separation in purely repulsive 
active systems \cite{cates2013active}. To the lowest order in density, they are 
typically of the form of $v(\varrho) = v_0(1 - a\phi \varrho)\, 
(v_0=\mathrm{const})$ which was also derived using linear response theory 
\cite{sharma2016communication}. The reason is simple: in regions with many 
particles, the effective swim speed must be reduced. It is interesting that we 
naturally obtain this form in the geometric integration procedure for hard 
interactions, and we can constitute that $a =2$ for steric interactions. 

\subsection{Hierarchy of hydrodynamic equations}
\label{section_hierarchy}

We now project the time-evolution equation for the full one-body PDF 
$p(\mathbf{x}, \theta, t)$ on its angular modes, where the zeroth-order mode 
is given by the mean particle PDF $\varrho(\mathbf{x}, t)$. This procedure 
generates a hierarchy of coupled particle differential equations for the time 
evolution of the modes. We analyse these equations to provide a footing to 
systematically close the hierarchy in the next Section.

\subsubsection{Expansion in harmonic modes.}

To overcome the angular dependence of the full one-body PDF, 
\textcolor{black}{we} expand the full one-body PDF in eigenfunctions of the 
rotation operator $\partial^2/\partial\theta^2$ in 
Eq.~\eref{final_full_one_body_equation}. This results in a Fourier expansion, 
in which modes of order $n \in\mathbb{N}_0$ have eigenvalue $-n^2$ in two 
dimensions. This expansion can be brought into the form
\begin{eqnarray}
\label{Cartesian_multipole_expansion_applied}
p(\mathbf{x}, \theta, t) = \textcolor{black}{\frac{1}{2\pi} \Big(} 
\varrho(\mathbf{x}, t) + \boldsymbol{\sigma}(\mathbf{x}, t) \cdot 
\hat{\mathbf{e}}(\theta) + \mathbf{Q}(\mathbf{x}, t) : 
\hat{\mathbf{e}}(\theta)\otimes\hat{\mathbf{e}}(\theta)   + \Upsilon(\theta) 
\textcolor{black}{\Big)},
\end{eqnarray}
where $\mathbf{Q} : \hat{\mathbf{e}}\otimes\hat{\mathbf{e}} = 
\textcolor{black}{Q_{\alpha\beta} \hat{e}_\beta\hat{e}_\alpha}$ denotes the 
full contraction with the outer product $\hat{\mathbf{e}}\otimes
\hat{\mathbf{e}}$, and $\Upsilon(\theta)$ refers to higher order modes. 
Notably, higher order (outer) products of the self-propulsion vector 
$\hat{\mathbf{e}}(\theta) = \left(\cos(\theta), \sin(\theta)\right)^\mathrm{T}$ 
naturally appear as modes in this expansion. Therefore, the expansion is also 
known as a Cartesian multipole expansion \cite{te2020relations} and finds wide 
use in the theory of liquid crystals \cite{degennes1995the}. Note that there 
exists an equivalent approach, \textcolor{black}{as} other works in active 
matter use an angular multipole expansion at this stage \cite{cates2013active, 
vuijk2021chemotaxis, muzzeddu2022active}, \textcolor{black}{where the second 
mode is replaced by $(\hat{\mathbf{e}}(\theta) \otimes 
\hat{\mathbf{e}}(\theta) - \mathbf{1}/2)$, where $\mathbf{1}$ denotes the 
identity tensor. The two expansions are equivalent, as $\mathbf{Q}$ is a 
traceless object, see below}. 

The coefficients of the expansion are given by the mean particle PDF 
$\varrho(\mathbf{x}, t)$ (zeroth-order mode), as defined above in 
Eq.~\eref{ad_hoc_definition_mean_density}, the mean polarisation 
$\boldsymbol{\sigma}(\mathbf{x}, t)$ (first-order mode), as defined above in 
Eq.~\eref{ad_hoc_definition_polarisation}, and the mean nematic order tensor 
(second-order mode)
\begin{equation}
\label{nematic_tensor}
     \mathbf{Q}(\mathbf{x}, t) = 4\int_0^{2\pi}\mathrm{d}\theta\ 
     p(\mathbf{x}, \theta, t)\ \left(\hat{\mathbf{e}}(\theta) \otimes 
     \hat{\mathbf{e}}(\theta) - \frac{\mathbf{1}}{2}\right).
 \end{equation}
Note that due to the similarity with hydrodynamic theories, the coefficients 
are sometimes referred to as \textit{hydrodynamic coefficients}, the hierarchy 
created by them as a \textit{hydrodynamic hierarchy} 
\cite{bertin2009hydrodynamic, cates2013active}. 

The inner product on the unit-sphere is defined 
 in the standard way as 
\begin{equation}
     \left\langle f(\theta), g(\theta)\right\rangle = 
     \int_0^{2\pi}\mathrm{d}\theta\ f(\theta) g(\theta),
\end{equation} 
for two functions $f$ and $g$. The modes of the Cartesian multipole 
expansion\textcolor{black}{, $\{\hat{\mathbf{e}}^n; n \in\mathbb{N}_0\}$,} 
form an orthogonal basis with respect to this scalar product. Therefore, 
we can project the full one-body PDF $p(\mathbf{x}, \theta, t)$ onto the 
modes and obtain the hydrodynamic coefficients 
\begin{numparts}
\begin{eqnarray}
\label{projection_p_on_density}
 \left\langle p(\mathbf{x}, \theta, t), 1\right\rangle = 
 \varrho(\mathbf{x}, t), \\
\label{projection_p_on_polarisation}
\left\langle p(\mathbf{x}, \theta, t), \hat{\mathbf{e}}(\theta)\right\rangle 
= \frac{1}{2}\ \boldsymbol{\sigma}(\mathbf{x}, t), \\
\label{projection_p_on_nematic}
\left\langle p(\mathbf{x}, \theta, t), \hat{\mathbf{e}}(\theta)\otimes 
\hat{\mathbf{e}}(\theta)\right\rangle = \frac{1}{4}\ \mathbf{Q}(\mathbf{x}, t).
\end{eqnarray}
\end{numparts}

\subsubsection{Time-evolution of modes.}

Using the orthogonality of modes, we can project the time-evolution 
equation for the full one-body PDF, Eq.~\eref{final_full_one_body_equation}, 
on each mode to obtain a time-evolution equation for the corresponding 
hydrodynamic coefficient. Projecting Eq.~\eref{final_full_one_body_equation} 
on the zeroth-order mode, we obtain a time-evolution equation of the mean 
particle PDF $\varrho$, which is given by
\begin{eqnarray}
\label{equation_for_mean_density}
\frac{\partial}{\partial t} \varrho(\mathbf{x},t)  &=  
D_T\ \nabla_\alpha\bigg[\left( 1 + 4\phi\ \varrho(\mathbf{x},t)\right)
\nabla_\alpha\varrho(\mathbf{x},t)\bigg] - \frac{v}{2}\ \nabla_\alpha 
\sigma_\alpha(\mathbf{x},t).
\end{eqnarray}
The mean particle PDF therefore obeys a continuity equation. This is expected 
on physical grounds as $\varrho$ is a conserved quantity. Specifically, this 
implies that $\varrho$ is a slow variable, i.e., a density perturbation of 
scale $\lambda$ relaxes on a time scale which diverges as $\lambda \to 
\infty$. This observation will be confronted with the time-evolution equation 
for the higher modes below. We further observe that the equation for the mean 
particle PDF needs an explicit input from the polarisation 
$\boldsymbol{\sigma}$. This coupling is induced by the activity $v$ in the 
model and persist for all modes, which generates a hierarchy. 
\textcolor{black}{Formally this hierarchy is similar to the 
famed BBGKY hierarchy of kinetic theory \cite{cercignani1994mathematical}, 
where also higher order modes implicitly alter the time-evolution of the 
mode in focus.}

Projecting Eq.~\eref{final_full_one_body_equation} onto the first order mode, 
we obtain a time-evolution equation for  the mean polarisation 
$\boldsymbol{\sigma}$, which is given in component-form by
\begin{eqnarray}
\label{equation_for_mean_polarisation}
    \frac{\partial}{\partial t} \sigma_\alpha(\mathbf{x},t) = D_T\ \nabla_\beta 
    \bigg[\left(1 - 2\phi\ \varrho(\mathbf{x},t)\right) \nabla_\beta 
    \sigma_\alpha(\mathbf{x},t) + 6\phi\ \sigma_\beta(\mathbf{x},t)\ 
    \nabla_\alpha \varrho(\mathbf{x},t)\bigg] \nonumber \\ 
    \quad - v\ \nabla_\beta \left[\left( 1 - 2\phi\ \varrho(\mathbf{x},t)\right) 
    \left(\frac{Q_{\beta\alpha}(\mathbf{x},t)}{2} +\varrho(\mathbf{x},t)\, 
    \delta_{\beta\alpha}\right)+ \phi\ \sigma_\beta(\mathbf{x},t)
    \sigma_\alpha(\mathbf{x},t)\right] \nonumber\\
    \quad -  D_R\ \Gamma_{\alpha\beta} \sigma_\beta(\mathbf{x},t).
\end{eqnarray}
Again, we observe that the time-evolution equation for the mean polarisation 
needs both input from the mean particle density $\varrho$ as well as from 
the nematic order tensor $\mathbf{Q}$. 
We further observe that in contrast to the time evolution of the mean particle 
density the structure of this equation is different. While $\varrho$ obeys a 
continuity equation, Eq.~\eref{equation_for_mean_polarisation} has a sink 
term: $- D_R\ \Gamma_{\alpha\beta} \sigma_\beta(\mathbf{x},t)$. The 
polarisation, and as we will see, all higher modes therefore are not 
conserved quantities, and their dynamics are governed by the associated time 
scale of the sink term. For the mean polarisation, this time scale is given 
by $\tau_1= 1/D_R$ \cite{sharma2017brownian, merlitz2018linear}.

This fundamental structural difference of the time-evolution equations 
arises in the $\theta$-term of the parental 
equation~\eref{final_full_one_body_equation} of the hierarchy. While for the 
polarisation we find that $\hat{\mathbf{e}}$ is the $n=1$st-order 
eigenfunction of the rotation operator with eigenvalue $-n^2 = -1$, for the 
zeroth-order mode $\varrho$, the eigenvalue is zero. For the chirality-induced 
term of Eq.~\eref{final_full_one_body_equation}, we observe a similar 
phenomenon. In the projection procedure, we find that
\begin{numparts}
\begin{eqnarray}
    \left\langle \hat{e}_\alpha, \omega \frac{\partial}{\partial\theta} 
    p\right\rangle &= \frac{\omega}{2\pi} \int_0^{2\pi}\mathrm{d}\theta\ 
    \hat{e}_\alpha(\theta)\ \frac{\partial}{\partial\theta} p(\mathbf{x}, 
    \theta, t) \\
    \label{periodicity_kills_boundary_2}
    \quad &= \frac{\omega}{2\pi} \hat{e}_\alpha\ p(\mathbf{x}, \theta, t) 
    \Big|_0^{2\pi} - \frac{\omega}{2\pi} \int_0^{2\pi}\mathrm{d}\theta\ 
    p(\mathbf{x}, \theta, t)\ \frac{\partial}{\partial\theta} 
    \hat{e}_\alpha(\theta) \\ 
    \quad &=  \varepsilon_{\alpha\beta}\ \frac{\omega}{2\pi}
    \int_0^{2\pi}\mathrm{d}\theta\ p(\mathbf{x}, \theta, t)\ 
    \hat{e}_\beta(\theta) \\
    \label{appereance_anti_symmetrie}
    \quad &= \frac{\omega}{2}\ \varepsilon_{\alpha\beta}\ 
    \sigma_\beta(\mathbf{x},t),
\end{eqnarray}
\end{numparts}
where we used that $\partial/\partial \theta\ \hat{e}_\alpha(\theta) 
= -\varepsilon_{\alpha\beta}\ \hat{e}_\beta(\theta)$ and 
$\boldsymbol{\varepsilon}$ is the Levi-\textcolor{black}{Civita} symbol in 
two dimensions\textcolor{black}{, defined by $\varepsilon_{xx}=\varepsilon_{yy} 
= 0$ and $\varepsilon_{xy} = -\varepsilon_{yx}=1$}. The boundary term in 
\eref{periodicity_kills_boundary_2} vanishes due to periodicity. When we are 
dealing with the zeroth-order mode, in contrast, the projection 
$\left\langle 1, \omega \frac{\partial}{\partial\theta} p\right\rangle$, 
reduces to the boundary term and therefore is zero due to periodicity 
in $\theta$.  

The joint effect of the rotation operator and the active chirality is the 
origin of the sink term $- D_R\ \Gamma_{\alpha\beta} 
\sigma_\beta(\mathbf{x},t)$, where $\boldsymbol{\Gamma}= 
(\mathbf{1} + \kappa \boldsymbol{\varepsilon})$ and $\kappa = \omega/D_R$ 
is the associated dimensionless parameter accounting for the effect of 
chirality. In a more general context, this parameter $\kappa$ is also 
referred to as the \textit{oddness} parameter \cite{hargus2021odd, 
han2021fluctuating, fruchart2023odd, kalz2024oscillatory}, since under a 
reversal of the direction of chirality, $\omega \to -\omega$, $\kappa$ changes 
sign and therefore switches the off-diagonal elements of the tensor 
$\boldsymbol{\Gamma} \to \boldsymbol{\Gamma}^\mathrm{T}$.

We find the time-evolution equation for the nematic order tensor from 
projecting Eq.~\eref{final_full_one_body_equation} on the second order 
mode $\hat{\mathbf{e}}\otimes\hat{\mathbf{e}}$
\begin{eqnarray}
\label{equation_for_nematic}
 \fl  \quad \frac{\partial}{\partial t} Q_{\alpha\beta}(\mathbf{x},t) &= 
 D_T\ \nabla_\gamma \bigg[(1 -2\phi\ \varrho(\mathbf{x},t))\ \nabla_\gamma 
 Q_{\alpha\beta}(\mathbf{x},t) + 6\phi\ Q_{\alpha\beta}(\mathbf{x},t)\  
 \nabla_\gamma \varrho(\mathbf{x},t))\bigg] \nonumber\\
 \fl   &\quad - v\ \nabla_\gamma \bigg[(1 - 2\phi\ \varrho(\mathbf{x},t))
 A_{\alpha\beta\gamma\delta}\  \sigma_\delta(\mathbf{x}, t) + \phi\ 
 \sigma_\gamma(\mathbf{x},t)  Q_{\alpha\beta}(\mathbf{x},t) + 
 \mathcal{O}(\Upsilon)\bigg] \nonumber \\
\fl   &\quad - 4D_R\ \tilde{\Gamma}_{\alpha\gamma} Q_{\gamma\beta}
(\mathbf{x},t). 
\end{eqnarray}
Here $A_{\alpha\beta\gamma\varepsilon} =\left(
\delta_{\alpha\gamma}\delta_{\beta\varepsilon} + 
\delta_{\alpha\varepsilon}\delta_{\beta\gamma} - 
\delta_{\alpha\beta}\delta_{\gamma\varepsilon}\right)/2$ and 
$\tilde{\boldsymbol{\Gamma}} = \left( \mathbf{1} + 
\kappa\, \boldsymbol{\varepsilon}/2\right)$ again accounts for the chirality. 
The structural similarities of this equation to the polarisation equation 
are that $(i)$ Eq.~\eref{equation_for_nematic} couples to neighbouring 
modes in the hierarchy and to the zeroth order mode $\varrho$ and $(ii)$ the 
relaxation dynamics of the nematic tensor is governed by the time scale 
$\tau_2 = 1/(4D_R)$ induced by the sink term of Eq.~\eref{equation_for_nematic}. 
Again we find that $\mathbf{Q}$, as all higher-order modes, is not conserved. 
We observe that the relaxation time scale originates in the eigenvalues of the 
rotation operator, i.e., the relaxation time scale of the $n$th mode is given 
by $\tau_n= \tau/n^2 = 1/(n^2 D_R)$, where we denote $\tau=\tau_1 = 1/D_R$ as 
the fundamental time scale and $n\geq1$.

In principle, the time-evolution equations for all higher-order modes can be 
found by the same projection procedure as presented before. This results in an 
infinite system of coupled time-evolution equations. To get an analytical, 
meaningful result from this hierarchy, we have to close this hierarchy based 
on physical arguments about negligible contributions from higher-order modes. 

\subsection{Closure of the hierarchy}
\label{section_closure_of_hierarchy}

We proceed to introduce the physical arguments to close the hierarchy by 
neglecting higher-order modes. Therefore, this problem naturally amounts to 
a perturbative analysis of the hierarchy.  As the result of the minimal 
non-trivial closure scheme, we find an \textcolor{black}{effective diffusion 
equation for} the mean particle density.

\subsubsection{Adiabatic approximation.}

It would be a formidable task to find a general solution for the $n$th order 
hierarchical equation and plug this into the $(n-1)$th order equation. By such 
a procedure, one aims to identify quasi-irrelevant modes for the effect on the 
time evolution of $\varrho$. But to our best knowledge, such a procedure has 
never been successfully applied before for a general order. Instead, it is 
convenient \cite{te2023derive, solon2015active, bertin2009hydrodynamic, 
vuijk2021chemotaxis, kreienkamp2022clustering, li2023towards, 
muzzeddu2023taxis, sevilla2016diffusion} to take advantage of the time scale 
separation in the system. The dynamics of the polarisation and all higher-order 
modes are governed by their respective time scale $\tau=\tau_1=1/D_R$ induced 
by rotational diffusion. $\tau$ then reasonably can be assumed to be much 
smaller than the relaxation time scale of the density, which can be arbitrarily 
large. Therefore, formally we will investigate the limit of $\tau \to 0$, 
called  \textit{adiabatic} or \textit{quasi-stationarity} approximation, 
since we adiabatically enslave the behaviour of higher order modes to the mean 
particle density and assume an instantaneous response to changes.

In some approaches, the adiabatic approximation is also handled differently. 
Especially in the context of phase separations, the nematic order tensor 
is assumed to be an adiabatic variable, but the polarisation is treated as a 
dynamical variable. This appears in active Brownian systems 
\cite{bertin2009hydrodynamic, yllanes2017many, ackerson1982correlations}, 
but also in active chiral systems \cite{kreienkamp2022clustering, 
liebchen2016pattern}. As explicitly shown there, such an ansatz can 
qualitatively improve the agreement of the analytical prediction with the 
numerical data, but it can rarely be treated fully analytically when aiming 
for the effect of higher-order modes on the relaxation of the mean particle 
PDF. In related hierarchy-closure problems, nevertheless, an analytic 
treatment of more than one dynamical variable and a shift of the adiabatic 
assumption to second order has recently been applied successfully in the 
framework of dynamical density functional theory \cite{tschopp2022first, 
tschopp2023superadiabatic}. Here many-body correlation functions form an 
analogous hierarchical problem, but the approach is analytically rather 
involved and it is not obvious how it can be applied to our situation. We, 
therefore, restrict our analysis to the basic situation and assume all 
higher-order modes to be adiabatic.

We will demonstrate the adiabatic approximation for the polarisation 
equation, but the same arguments hold true for all higher-order modes. 
Pointing out the essentials, Eq.~\eref{equation_for_mean_polarisation} can 
be written as 
\begin{equation}
    \frac{\partial}{\partial t} \sigma_\alpha + \frac{1}{\tau} 
    \Gamma_{\alpha\beta}  \sigma_\beta = f_\alpha(\varrho, 
    \boldsymbol{\sigma}, \mathbf{Q}), 
\end{equation}
where $\mathbf{f}$ denotes the leftover gradient-structure terms of 
Eq.~\eref{equation_for_mean_polarisation}. We can solve this equation 
formally by
\begin{eqnarray}
\label{formal_solution_quasi_stationarity}
    \sigma_\alpha(\mathbf{x},t) = \mathrm{e}^{-\Gamma_{\alpha\beta} 
    t/\tau}\sigma_\beta(\mathbf{x},0) + \int_0^t\mathrm{d}t^\prime\ 
    \mathrm{e}^{-\Gamma_{\alpha\beta} |t-t^\prime|/\tau} 
    f_\beta(\varrho, \boldsymbol{\sigma}, \mathbf{Q}), 
\end{eqnarray}
where $\varrho, \boldsymbol{\sigma}$ and $\mathbf{Q}$ have to be evaluated 
at $t^\prime$ inside the integral. The integrating factor 
$\mathrm{e}^{-\Gamma_{\alpha\beta} t/\tau}$ is defined as the usual 
matrix-exponential and can be reformulated as 
\begin{eqnarray}
    \mathrm{e}^{-\Gamma_{\alpha\beta} t/\tau} 
    &= \mathrm{e}^{- t/\tau}\ \left(\cos\left(\kappa\ \frac{t}{\tau}\right) 
    \delta_{\alpha\beta} - \sin\left(\kappa\ \frac{t}{\tau}\right) 
    \varepsilon_{\alpha\beta} \right),
\end{eqnarray}
using the anti-symmetry property of $\boldsymbol{\Gamma}$. The active 
chirality \textcolor{black}{$(\kappa \propto \omega)$} therefore results 
in oscillations, which decay exponentially on the typical time scale $\tau$. 
The exponential factor further allows us to write $\int_0^\infty\mathrm{d}t\ 
\mathrm{exp}(-\Gamma_{\alpha\beta} t/\tau) = \tau \Gamma_{\alpha\beta}^{-1}$, 
which together with $\lim_{\tau \to 0} \mathrm{exp}(-\Gamma_{\alpha\beta} 
|t|/\tau) = 0_{\alpha\beta}$ for $|t| >0$ results in $\lim_{\tau \to 0} 
\Gamma_{\alpha\gamma}\ \mathrm{exp}(-\Gamma_{\gamma\beta} |t|/\tau) = 2\tau\ 
\delta_{\alpha\beta} \delta(t)$. Here
the factor $2$ originates from the integrals taken to be for positive 
times only. The formal solution of Eq.~\eref{formal_solution_quasi_stationarity} 
evaluated in the limit of $\tau \to 0$ thus reads
\begin{equation}
    \lim_{\tau\to 0} \sigma_\alpha(\mathbf{x},t) = \tau 
    \Gamma^{-1}_{\alpha\beta}\ f_\beta(\varrho, \boldsymbol{\sigma}, 
    \mathbf{Q}), 
\end{equation}
where now $\varrho, \boldsymbol{\sigma}$ and $\mathbf{Q}$ are functions of $t$.
The corresponding adiabatic approximation for the nematic order tensor similarly 
reads
\begin{equation}
\lim_{\tau\to 0}  Q_{\alpha\beta}(\mathbf{x},t) = 4\tau 
\tilde{\Gamma}^{-1}_{\alpha\gamma}\  g_{\gamma\beta}(\varrho, \boldsymbol{\sigma}, 
\mathbf{Q}, \Upsilon),
\end{equation}
where $\mathbf{g}$ accounts for the leftover terms of 
Eq.~\eref{equation_for_nematic} and the modes again are evaluated at $t$. 

\subsubsection{Closure of the hierarchy via perturbation approach.}

The adiabatic approximation as such does not suffice to close the hierarchy. 
The coupling to higher order modes in the hierarchy for each mode enters via 
the activity-induced term $-v\ \nabla \cdot \left[( 1 - 2\phi\ \varrho)\ 
\hat{\mathbf{e}} p + 2\phi\ \boldsymbol{\sigma} p \right]$ from the parental 
equation~\eref{final_full_one_body_equation}. This coupling already takes 
place in interaction-free considerations ($\phi = 0$), as it specifically 
arises due to the appearance of the self-propulsion vector 
\textcolor{black}{and first-order mode} $\hat{\mathbf{e}}$. To effectively 
close the hierarchy we therefore have to argue that higher-order modes can be 
neglected as compared to lower ones. This consequently turns the closure of 
the hierarchy into a perturbation problem. 

We introduce a dimensionless time via the natural time scale of the system 
$\tau$ and dimensionless space via the mean-particle distance 
$l_\mathrm{dist} = L \sqrt{\phi}$. The two natural physical length scales 
in the system are the persistence length $l_\mathrm{pers} = v/D_R$ induced 
by activity and the diffusion length scale $l_\mathrm{diff} = \sqrt{D_T/D_R}$ 
induced by thermal equilibrium fluctuations \cite{bruna2022phase}. We denote 
the parameters, which arise when comparing the physical length scales with the 
dimensionless length scale $l_\mathrm{dist}$ as $\varepsilon_p$ and 
$\varepsilon_D$ and observe the following relation between them:
\begin{eqnarray}
\label{perturbation_parameters}
    \varepsilon_p = \frac{l_\mathrm{pers}}{l_\mathrm{dist}}, \quad  
    \varepsilon_D = \frac{l_\mathrm{diff}}{l_\mathrm{dist}}, \quad 
    \mathrm{and}\quad \frac{\varepsilon_p}{\varepsilon_D} = 
    \frac{l_\mathrm{pers}}{l_\mathrm{diff}} = \frac{v}{\sqrt{D_T D_R}} =
    \mathrm{Pe}.
\end{eqnarray}
Here Pe stands for the P{\'e}clet number, which measures the relation of 
activity-induced self-propulsion versus thermally induced displacement. 

Here we take  $\varepsilon_p$ and $\varepsilon_D$ to be two independent 
perturbation parameters. We show that depending on their relation to the 
P{\'e}clet number, different field theoretic descriptions can be obtained. 
The adiabatic assumption thereby justifies independent small active and passive 
length scales, since both $l_\mathrm{pers} \propto \tau$ and $l_\mathrm{diff} 
\propto \tau$. This treatment is in contrast to previous work, where this 
subtle point of taking both parameters small and not only their ratio (the 
P{\'e}clet number) was often left implicit or was overlooked. 

$\varepsilon_p$ and $\varepsilon_D$ compare the active and passive physical 
length scales in the system to the dimensionless length scale, which we choose 
as the mean particle-particle distance. This is a careful choice, since the 
other natural length scales in the system, the particle diameter $d$ 
(which also equals the interaction length scale for hard systems) and the 
typical box size $L$ were too small and too big, respectively, and already 
form the small parameter $\phi \propto (d/L)^2$. \textcolor{black}{Together} 
with the assumption of a dilute system, the mean particle-particle distance 
appears as the correct length scale interpolating between a too-narrow or 
too-coarse-grained view on the dynamics.

The aforementioned analysis reveals that the coupling to higher-order modes 
takes place in the activity-induced term at each order. To close the 
hierarchy, we therefore need to decide up to which order we consider the 
parameter $\varepsilon_p$. For this work, and referring to what is typical 
in the literature \cite{te2023derive, solon2015active, bertin2009hydrodynamic, 
vuijk2021chemotaxis, kreienkamp2022clustering, li2023towards, 
muzzeddu2023taxis}, we truncate the hierarchy at order 
$\mathcal{O}(\varepsilon_p^3)$. We thus ignore contributions from higher 
modes such as $\Upsilon$ in the nematic equation. Together with the adiabatic 
assumption Eq.~\eref{equation_for_nematic} for the nematic tensor thus becomes
\begin{eqnarray}
\label{small_activity_nematic}
Q_{\alpha\beta}(\mathbf{x},t) &= \frac{\varepsilon_D}{4} 
\tilde{\Gamma}^{-1}_{\alpha\delta}\nabla_\gamma \bigg[(1 - 2\phi\ 
\varrho(\mathbf{x},t))\ \nabla_\gamma Q_{\delta\beta}(\mathbf{x},t) \nonumber \\
&\quad + 6\phi\ Q_{\delta\beta}(\mathbf{x},t)\ \nabla_\gamma 
\varrho(\mathbf{x},t))\bigg]. 
\end{eqnarray}

Eq.~\eref{small_activity_nematic} constitutes a fixed-point problem for 
$\mathbf{Q}$. We observe that Eq.~\eref{small_activity_nematic} has no sink 
term and therefore has a definite, perturbation-free solution given by 
$\mathbf{Q} = \mathbf{0}$. This is a generic observation: closing the hierarchy at 
order $\mathcal{O}(\varepsilon_p^n)$ results in the fixed-point problem for 
the $(n-1)$th order mode to only have the trivial solution.

\subsubsection{Effective diffusion equation.}

In the following, we investigate the resulting time-evolution equations for 
the density for different orders of truncation in the polarisation equation. 
We start with the first non-trivial case by allowing for terms of order 
$\mathcal{O}(\varepsilon_p^1)$ for the polarisation.
Here, the adiabatic polarisation reads 
\begin{eqnarray}
\label{sigma_model_B}
    \sigma_\alpha = -\varepsilon_p\ \Gamma^{-1}_{\alpha\beta} 
    \bigg[\left( 1 - 4\phi\ \varrho \right)\nabla_\beta\varrho \bigg].
\end{eqnarray}
Note that since the diffusive parameter $\varepsilon_D$ originates from a 
Laplace operator it only appears at even powers. Thus, there is no term at 
order $\mathcal{O}(\textcolor{black}{\varepsilon_p}\varepsilon_D)$ that could 
be included at this closure of the polarisation. The mean particle density 
from Eq.~\eref{equation_for_mean_density} at this order becomes
\begin{eqnarray}
\label{model_B_pre_constant}
    \frac{\partial}{\partial t} \varrho(\mathbf{x},t)  &=  \varepsilon_D^2\ 
    \nabla_\alpha\bigg[\left( 1 + 4\phi\ \varrho(\mathbf{x},t)\right)\ 
    \nabla_\alpha\varrho(\mathbf{x},t)\bigg] \nonumber\\ 
    &\quad + \frac{\varepsilon_p^2}{2} \frac{1}{1 + \kappa^2}\  
    \nabla_\alpha\bigg[\left( 1 - 4\phi\ \varrho(\mathbf{x},t)\right)\ 
    \nabla_\alpha\varrho(\mathbf{x},t)\bigg].
\end{eqnarray}

As apparent, the chosen closure scheme results in a time-evolution 
equation of the mean particle density at order $\mathcal{O}(\varepsilon_p^2)$ 
and $\mathcal{O}(\varepsilon_D^2)$. Now, by comparing the terms under 
consideration with the truncated terms we can learn about the regime of 
validity of Eq.~\eref{model_B_pre_constant}. This gives us that 
$(i)$ $\varepsilon_p^2 \gg \varepsilon_p^3$, which is consistent with our 
perturbative assumption of $\varepsilon_p\ll 1$, but we also find that 
$(ii)$ $\varepsilon_D^2 \gg \varepsilon_p^3$, which tells us that 
Eq.~\eref{model_B_pre_constant} is valid in the regime of 
$\mathrm{Pe} \ll 1/\varepsilon_p^{1/2}$. Together $(i)$ and $(ii)$ do not 
form a precise upper bound to the P{\'e}clet number,\ \textcolor{black}{in fact 
it can be arbitrarily large} and therefore an analysis of the microscopic 
parameters is essential when using this field-theoretical description.

Eq.~\eref{model_B_pre_constant} is written in dimensionless form. Thus we 
can reintroduce the units of space and time, i.e., $\tau$ for time and 
$l_\mathrm{dist}$ for space, compare also relation 
\eref{perturbation_parameters}. The time-evolution equation for the mean 
density becomes
\begin{eqnarray}
\label{model_B}
    \frac{\partial}{\partial t} \varrho(\mathbf{x},t) = 
    \nabla\cdot\bigg[\left(D_T^\mathrm{eff}(\varrho) + 
    D_A^\mathrm{eff}(\varrho) \right)\nabla \varrho(\mathbf{x},t)\bigg]. 
\end{eqnarray}
The time evolution of $\varrho$ at this order thus follows 
\textcolor{black}{an effective diffusion equation~\cite{bialke2013microscopic, 
cates2013active,bickmann2022analytical}, where } $D_T^\mathrm{eff} + 
D_A^\mathrm{eff}$ form the interaction-corrected diffusion coefficients 
due to thermal and active motion, respectively, and are given by
\begin{numparts}
\begin{eqnarray}
\label{model_B_thermal_diffusion_coefficient}
    D_T^\mathrm{eff}(\varrho) &= D_T \left(1 + 4\phi\ \varrho(\mathbf{x},t)
    \right), \\
    \label{model_B_active_diffusion_coefficient}
    D_A^\mathrm{eff}(\varrho) &= \textcolor{black}{D_A^\omega} \left(1 - 
    4\phi\ \varrho(\mathbf{x},t)\right).
\end{eqnarray}
\end{numparts}
Here $D_T$ stands for the thermal (equilibrium) diffusion coefficient, as 
introduced in the microscopic Langevin description 
\eref{lanegvin_poistion_setup_acp}. $\textcolor{black}{D_A^\omega = 
D_A^0 / (1 + \kappa^2)} = v^2/( 2D_R (1 + \kappa^2))$ is the 
\textcolor{black}{chirality-affected} active diffusion coefficient, which 
describes the ballistic motion. \textcolor{black}{$D_A^0$ is thereby a 
characteristic} of a purely active particle and relates it to the randomisation 
of the self-propulsion vector due to rotational diffusion. 
\textcolor{black}{In accordance with observations in the literature  
\cite{bickmann2022analytical, nourhani2013chiral,chan2024chiral, 
van2019interparticle}, active chirality $\omega$ rescales the active diffusion 
$D_A^0$, as $D_A^\omega$ can also be written as} 
$\textcolor{black}{D_A^\omega} = v^2 D_R/(2(\omega^2 + D_R^2))$. Remember 
that $\kappa = \omega/D_R$.

Relations \eref{model_B_thermal_diffusion_coefficient} and 
\eref{model_B_active_diffusion_coefficient} state that hard-core interactions 
on the one hand enhance the thermal diffusion and on the other hand reduce the 
active diffusion. For the passive motion, this is in accordance with the 
observation that the collective diffusion, in what sense 
$D_T^\mathrm{eff}(\varrho)$ also can be interpreted, gets enhanced by 
steric interactions \cite{kalz2022collisions, kalz2022diffusion, 
dhont1996introduction}. 
For the active motion similarly the interaction-reduction of the associated 
diffusion coefficient $D_A^\mathrm{eff}$ is no surprise. It rather can be 
regarded as an analytic necessity for the strong theory-, simulation-, and 
experiment-supported existence of motility-induced phase separations 
\cite{fily2012athermal, buttinoni2013dynamical, redner2013structure, 
liu2013phase, wysocki2014cooperative}, i.e., the phenomenon that purely 
repulsive active systems can phase-separate as a function of particle density. 
This phenomenon is only possible if the associated diffusion process becomes 
unstable, i.e., the effective diffusion coefficient can formally turn negative. 

\subsection{Active Model B +}
\label{section_AMB_plus}

In this Section, we go one step beyond the simplest non-trivial closure 
approximation by considering mixed terms of the perturbation parameters. 
This amounts to a much richer field-theoretical description but for the price 
of a much narrower regime of validity. We find that the dynamics of the mean 
particle PDF follow the recently introduced AMB+, to which we, therefore have 
first-principles access for the parameters. Surprisingly here we find that the 
characteristic AMB+ parameters all change sign as a function of chirality. 

\subsubsection{Effect of mixed perturbation parameters.}

In this work, we treat both the activity as well as the thermal 
diffusion-induced length scales as small compared to the inter-particle 
distance. This is justified by the assumption of a dilute system. We are 
thus formally dealing with a two-parameter perturbation theory. It is 
therefore natural that besides arguing for the smallness of each of the 
parameters $\varepsilon_D \ll 1$ and $\varepsilon_p\ll 1$, we also have to 
specify their relative size. This is only possible by making a third assumption 
about the P{\'e}clet number, the natural scale relating the active and passive 
motion of a tracer. 

We already observed from Eq.~\eref{sigma_model_B} that the diffusion 
length-scale parameter $\varepsilon_D$ only appears in even powers (due to its 
origin in the second order Laplace operator). Therefore, we can go a step 
further in the truncation scheme, by considering the polarisation up to order $\mathcal{O}(\textcolor{black}
{\varepsilon_p^1}\varepsilon_D^2)$. The adiabatic polarisation can 
self-consistently be obtained from Eq.~\eref{equation_for_mean_polarisation} 
at the desired order to be
\begin{eqnarray}
    \sigma_\alpha &= -\varepsilon_p\, \Gamma^{-1}_{\alpha\beta} 
    \bigg[\left( 1 - 4\phi\ \varrho \right)\ \nabla_\beta\varrho \bigg] 
    \nonumber \\ 
    &\quad -\varepsilon_p\varepsilon_D^2\, 
    \left(\Gamma^{-1}\right)^2_{\alpha\beta} \bigg[\nabla_\gamma^2 - 
    2\phi\left(3\varrho\ \nabla_\gamma^2 - (\nabla_\gamma^2\varrho) + 
    2(\nabla_\gamma \varrho)\ \nabla_\gamma\right)\bigg] \nabla_\beta\varrho.
\end{eqnarray}
We observe here that the higher-order gradient terms of the mean particle 
density arise together with a matrix product of the chiral matrix 
$\boldsymbol{\Gamma}$. For an arbitrary vector $\mathbf{a}$, the contraction 
of the relevant matrices contributes
\begin{numparts}
\begin{eqnarray}
    a_\alpha\ \Gamma^{-1}_{\alpha\beta}\ a_\beta = \frac{1}{1 + \kappa^2}\ 
    a_\alpha a_\alpha,\\ 
    a_\alpha\ \left(\Gamma^{-1}\right)^2_{\alpha\beta}\ a_\beta = 
    \frac{1 - \kappa^2}{(1 + \kappa^2)^2}\ a_\alpha a_\alpha, 
\end{eqnarray}
\end{numparts}
where due to the antisymmetric structure of $\boldsymbol{\Gamma}^{-1} = 
(\mathbf{1} -\kappa\boldsymbol{\varepsilon})/(1 + \kappa^2)$ only the 
respective diagonal elements are relevant in the full contraction. 

Inserting this expression for the polarisation into 
Eq.~\eref{equation_for_mean_density} for the time-evolution of the mean 
particle density $\varrho$, we obtain 
\begin{eqnarray}
\label{FP_pre_AMB_plus}
\frac{\partial}{\partial t} \varrho(\mathbf{x},t)  = \varepsilon_D^2 
\nabla_\alpha\bigg[\left( 1 + 4\phi\ \varrho\right) 
\nabla_\alpha\varrho\bigg]  + \frac{\varepsilon_p^2}{2} 
\frac{1}{1 + \kappa^2}\nabla_\alpha \bigg[\left( 1 - 4\phi\ \varrho \right) 
\nabla_\alpha\varrho \bigg] \nonumber \\
\quad + \frac{\varepsilon_p^2\varepsilon_D^2}{2} 
\frac{1- \kappa^2}{(1 + \kappa^2)^2}\nabla_\alpha \bigg[\nabla_\gamma^2 - 
2\phi\left(3\varrho \nabla_\gamma^2 - (\nabla_\gamma^2\varrho) + 
2(\nabla_\gamma \varrho)\nabla_\gamma\right)\bigg] \nabla_\alpha\varrho.
\end{eqnarray}
It is now obvious that, to obtain this field-theoretical description for the 
density compared to Eq.~\eref{model_B_pre_constant}, we further encounter the 
mixed term of order $\mathcal{O}(\varepsilon_p^2\varepsilon_D^2)$, which is 
only possible if this term is assumed to be much greater then the disregarded 
terme, e.g. that of order $\mathcal{O}(\varepsilon_p^3)$. An easy algebraic analysis shows 
that this is only possible when $\mathrm{Pe} \ll \varepsilon_D \ll 1$ due to 
the perturbative closure scheme. 

\subsubsection{Active Model B +.}

If we reintroduce physical units for space and time, i.e.,  $\tau$ for the 
time $l_\mathrm{dist}$ for the space, compare also relation 
\eref{perturbation_parameters}, we can arrange Eq.~\eref{FP_pre_AMB_plus} 
in the form 
\begin{eqnarray}
\label{AMB_plus_coefficients}
    \frac{\partial}{\partial t} \varrho(\mathbf{x},t) &= a \nabla^2\varrho + 
    b \nabla^2(\varrho^2) - k_0 \nabla^4\varrho -k_1\left[
        \nabla^2(\nabla\varrho)^2 +2 \nabla^2(\varrho\nabla^2\varrho)\right]
        \nonumber\\
    &\quad+ \lambda \nabla^2\left(\nabla\varrho\right)^2 - \xi\nabla\cdot
    (\nabla\varrho)(\nabla^2\varrho),
\end{eqnarray}
where 
\begin{numparts}
    \label{AMB_plus_coeficients}
\begin{eqnarray}
\label{AMB_plus_coefficient_a}
    a = D_T + \textcolor{black}{D_A^\omega}, \\
    b = 2\phi \left(D_T - \textcolor{black}{D_A^\omega}\right), \\
    \label{AMB_plus_coefficient_lambda}
   \lambda = \phi \textcolor{black}{D_A^\omega} D_T\ \frac{1 - \kappa^2}
   {1 + \kappa^2}, \\
    \label{AMB_plus_coefficient_xi}
    \xi = -8\phi \textcolor{black}{D_A^\omega} D_T\ \frac{1 - \kappa^2}
    {1 + \kappa^2}, \\
    \label{AMB_plus_coefficient_k}
    k\textcolor{black}{[}\varrho\textcolor{black}{]} = \textcolor{black}
    {D_A^\omega} D_T\ \frac{1 - \kappa^2}{1 + \kappa^2} \left(-1 + 6\phi\ 
    \varrho(\mathbf{x},t)\right),
\end{eqnarray}
\end{numparts}
and $k\textcolor{black}{[}\varrho\textcolor{black}{]} = k_0 + 2k_1\ 
\varrho(\mathbf{x},t)$. Eq.~\eref{AMB_plus_coefficients} is known as the 
(deterministic) AMB+ \cite{tjhung2018cluster, cates2023classical, 
nardini2017entropy,  caballero2018bulk}, and can be rearranged into the form
\begin{numparts}
\label{full_AMB_plus}
\begin{eqnarray}
\label{AMB_plus_gradient_form}
    \frac{\partial}{\partial t} \varrho(\mathbf{x},t) &= -\nabla 
    \cdot\left[ -\nabla \left(\frac{\delta \mathcal{F}}{\delta \varrho} + 
    \lambda (\nabla \varrho)^2\right) + \xi (\nabla\varrho)(\nabla^2\varrho)
    \right],
\end{eqnarray}
where 
\begin{eqnarray}
\label{AMB_plus_free_energy}
    \mathcal{F}[\varrho] = \int\mathrm{d}\mathbf{r} \left(f_0
    \textcolor{black}{[}\varrho\textcolor{black}{]} + 
    \frac{k\textcolor{black}{[}\varrho\textcolor{black}{]}}{2} 
    (\nabla\varrho)^2 \right)
\end{eqnarray}
\end{numparts}
is the free-energy functional and $f_0\textcolor{black}{[}
\varrho\textcolor{black}{]} = a/2\ \varrho^2 + b/3\ \varrho^3$ is the bulk 
free-energy density. Note that typically $\varrho^\prime = \varrho - 
\varrho_\mathrm{MF}$ is taken to be the order parameter, where 
$\varrho_\mathrm{MF}$ is the mean-field critical value of the density. 
If one does so, a term $\propto \varrho^3$ is forbidden by symmetry in $f_0$, 
but we instead work with the density $\varrho$ itself. Note that
Eq.~\eref{full_AMB_plus} was first written down based on phenomenologically 
accounting for systems with broken detailed-balance to lowest order terms 
in Refs.~\textcolor{black}{\cite{tjhung2018cluster, nardini2017entropy}}.

\subsubsection{First-principles expressions for field-theoretical parameters.}

The parameter $k\textcolor{black}{[}\varrho\textcolor{black}{]}$ in the free 
energy is known as the Cahn-Hilliard parameter \cite{cahn1958free}. When 
first introduced, this parameter was the minimal attempt in an equilibrium 
model to extend the bulk free energy $\mathcal{F}$ to further include 
density-gradient-induced inhomogeneities into the description. Via this 
very successful approach, field theories could nicely capture 
phase-separation dynamics in equilibrium models \cite{bray2002theory, 
desai2009dynamics}. In the general formulation, 
$k\textcolor{black}{[}\varrho\textcolor{black}{]}$ is density-dependent, and a 
Taylor expansion to lowest order in the density gives 
$k\textcolor{black}{[}\varrho\textcolor{black}{]} = k_0 + 2 k_1 \varrho$. 
The coefficients $k_0$ and $k_1$ are found in our model by comparing the 
first-principles time-evolution equation~\eref{FP_pre_AMB_plus} with the 
phenomenological equation of the AMB+ \eref{AMB_plus_coefficients}. From an 
equilibrium perspective, it might appear surprising that from 
Eq.~\eref{AMB_plus_coefficient_k} we find that $k_0 < 0$ for a non-chiral 
system ($\kappa=0$) and thus also $k\textcolor{black}{[}
\varrho\textcolor{black}{]} <0$ to lowest order. In an equilibrium field theory, 
this would lead to an unbounded free energy and the non-physical possibility 
of a system minimising its free energy by creating more and more interfaces due 
to phase separation. But for an active field theory such as the AMB+, the time 
evolution is not solely governed by a free-energy structure. Further terms 
$\propto \lambda, \xi$ balance the free-energy governed evolution, and thus 
equilibrium intuition can fail. Note also that in our first-principles 
derivation for active systems $k\textcolor{black}{[}\varrho\textcolor{black}{]} 
\propto \textcolor{black}{D_A^\omega}$, and therefore this quantity vanishes 
as the model is approaching an equilibrium situation 
($\textcolor{black}{D_A^\omega} \to 0$). Finally, the observation of 
$k_0 < 0$ is consistent with other works on the AMB+ \cite{tjhung2018cluster, 
te2023derive}, but (active) chirality adds a new perspective, since here 
$k_0$ and $k_1$ change sign as $\kappa>1$. 

\begin{figure}
    \centering
    \includegraphics[width=0.6\columnwidth]{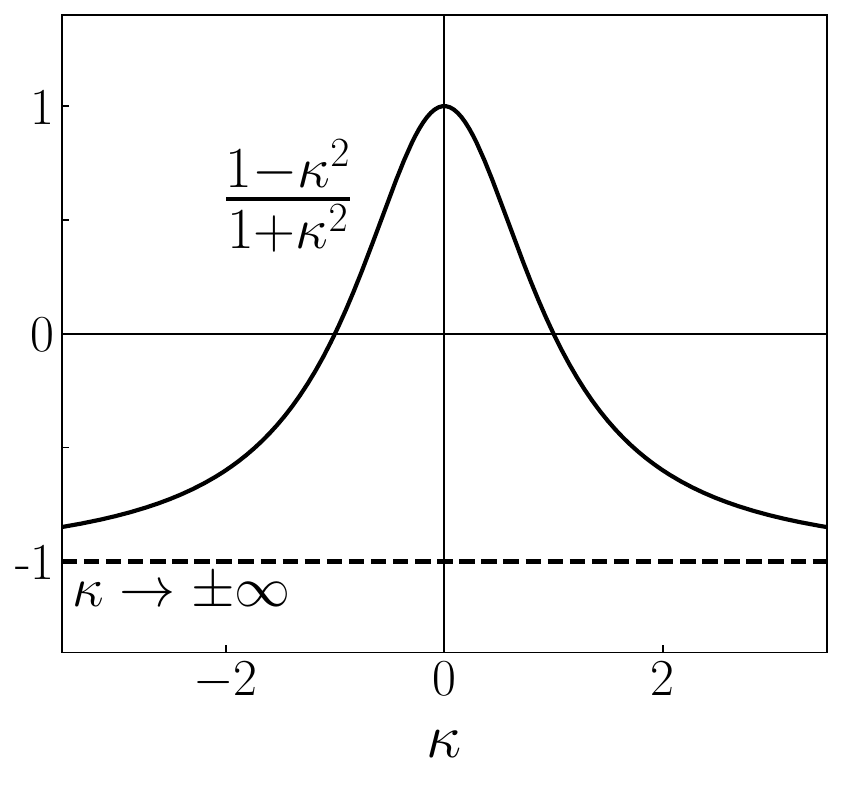}
    \caption{For ACPs one can define the parameter $\kappa = \omega/D_R$, which 
    is a measure of active chirality $\omega$ versus rotational diffusion $D_R$ 
    of the particle. In a related context, this parameter is known as the 
    \textit{oddness} parameter \cite{kalz2022diffusion,  muzzeddu2022active, 
    hargus2021odd, kalz2024oscillatory}. The characteristic parameters of the 
    AMB+, $\lambda, -\xi$ and $k\textcolor{black}{[}\varrho\textcolor{black}{]}$ 
    all scale $\propto (1 - \kappa^2)/(1 + \kappa^2)$, see also 
    Eqs.~\eref{AMB_plus_coefficient_lambda} to \eref{AMB_plus_coefficient_k}. 
    Thus, 
    they are positive as long as $|\omega| < D_R$, i.e., the circular motion is 
    dominated by diffusion, and are negative for $|\omega| > D_R$, i.e., the 
    motion is dominated by activity. Note that the sign of $\omega$ accounts 
    for the clockwise or counterclockwise direction of the active chirality.}
    \label{fig:kappa_curve}
\end{figure}

The observation, that the coefficients which are induced by activity and 
interactions ($\lambda, \xi, k\textcolor{black}{[}\varrho\textcolor{black}{]} 
\propto \phi \textcolor{black}{D_A^\omega}$) can change their sign as a 
function of chirality is rather surprising. From the Langevin dynamics and the 
integration procedures, we would have not expected that chirality and 
interactions could interplay such, that they lead to physical consequences. 
In the hard-core interacting scenario the angular coordinate is not altered 
by the excluded-volume interaction of the spatial coordinate, and therefore 
chirality does not affect the time-evolution of the full one-body PDF beyond 
the (trivial) interaction-free terms, as can be seen from 
Eq.~\eref{final_full_one_body_equation}. Any interplay of spatial and angular 
\textcolor{black}{coordinates} is only introduced in the projection on the 
hydrodynamic modes. We can nevertheless qualitatively 
justify why the change in the behaviour happens at $\kappa=1$. Similar to the 
rotational diffusion coefficient, which introduces a time scale in the 
system $\tau = \tau_\mathrm{diff} = 1/D_R$, the active chirality introduces 
a time scale as well, namely, $\tau_\mathrm{act} = 1/\omega$. These time scales 
represent the noise and the deterministic circular contribution to the active 
motion of the particle, respectively. Hence the parameter $\ \kappa=\omega/D_R 
= \tau_\mathrm{diff}/\tau_\mathrm{act}$ measures which contribution is 
dominating the motion of a particle, similar to what was recently reported 
in Ref. \textcolor{black}{\cite{chan2024chiral, liao2018clustering}}. That 
means that $\kappa<1$ corresponds to a system, in which the diffusive motion 
dominates the active particle, whereas $\kappa>1$ corresponds to a 
deterministic-circular-motion determined motion, see also 
Fig.~\ref{fig:kappa_curve}. Interestingly, when both effects are of equal 
magnitude, $\lambda, \xi, k\textcolor{black}{[}\varrho\textcolor{black}{]} =0$, 
the system again behaves as described by the 
\textcolor{black}{effective diffusion} dynamics of Eq.~\eref{model_B}.

The phenomenological parameters $\lambda$ and $\xi$ represent the active 
generalisations of equilibrium Model B \cite{hohenberg1977theory} 
($\partial\varrho/\partial t = -\nabla \mathbf{J}_\mathrm{eq}= \nabla 
(\delta \mathcal{F}[\varrho]/ \delta \varrho)$, where $\varrho$ is the 
conserved order parameter, $\mathbf{J}_\mathrm{eq}$ the equilibrium 
(deterministic) current and $\mathcal{F}$ is the equilibrium free energy). 
Since this model was constructed on the basic principle of the system to obey 
detailed-balance, to describe active matter, this restriction had to be 
overcome. A first step in the development of active field theories was the 
so-called \textit{Active Model B (AMB)}  \cite{stenhammar2013continuum, 
wittkowski2014scalar}, where $\lambda \neq 0$ but $\xi=0$. This model 
still attracts interest as it represents the first step towards an active 
field theory \cite{zakine2023unveiling, o2023nonequilibrium}, but it was 
shown that even though quantitatively the coexisting liquid and vapour densities 
are changed, the AMB cannot report any quantitative changes in the coarsening 
dynamics as compared to the known dynamics found by phenomenologically 
introducing activity in the Model B \cite{wittkowski2014scalar}. Note that 
Eq.~\eref{model_B} as the lower order result of the closure scheme, has such 
a Model B structure with coefficients altered by activity.

The sought qualitative changes in the phase-separation dynamics of active 
matter could thereafter be found by the generalised AMB+ where $\lambda \neq 0$ 
and additionally $\xi\neq0$. Similarly to the AMB, the AMB+ as the most 
general isotropic model at this order goes beyond the typical free-energy structure 
of Model B ($\lambda\neq 0$). For the AMB, the $\lambda$-term defines a 
(local) non-equilibrium chemical potential $\mu_\mathrm{neq} = \delta 
\mathcal{F}/\delta \varrho + \lambda (\nabla\varrho)^2$, since the 
current is still of a gradient form. For the AMB+ instead the $\xi$-induced 
current cannot be put into a gradient structure anymore and therefore allows 
for circulating real-space currents $\nabla \wedge \mathbf{J}_\mathrm{neq} 
\propto \xi$. Further, as a result of the non-gradient structure of the 
current, the non-equilibrium chemical potential becomes non-local 
\cite{tjhung2018cluster, cates2023classical} and therefore amounts to 
fundamental differences of AMB and AMB+. As these studies suggest, this 
non-locality seems to be a necessary ingredient to describe the behaviour 
of active matter from a field theoretical perspective.

\textcolor{black}{Finally, it is} interesting to note that as a result of our 
first principles approach, the AMB+ appears as the most natural choice of an 
active field theory which relies on an expansion in terms of density gradients. 
We observe from Eqs.~\eref{AMB_plus_coefficient_lambda} and 
\eref{AMB_plus_coefficient_xi} that $\lambda = -8 \xi$. Thus, it is not 
reasonable to include one but leave out the other parameter in the field-
theoretical description, \textcolor{black}{similarly to what was pointed out 
recently \cite{rapp2019systematic}}. We further observe that, as argued 
phenomenologically, 
both $\lambda, \xi \propto D_T \textcolor{black}{D_A^\omega}$, such that the 
AMB+ reduces to \textcolor{black}{an interaction-corrected, passive diffusion 
equation} for $\textcolor{black}{D_A^\omega}\to 0$. Interestingly, 
it also reduces to the non-equilibrium effective diffusion equation 
\eref{model_B} as $D_T \to 0$, i.e., when the 
thermal motion becomes negligible compared to the activity. Lastly, as expected, 
$\lambda$ and $\xi$ are only present in an interacting system 
($\lambda, \xi \propto \phi$), since they are known to alter the 
phase-separation dynamics. 

\begin{figure}
    \centering
    \includegraphics[clip, trim = 1.5cm 0cm 2cm 0cm, width=\columnwidth]
    {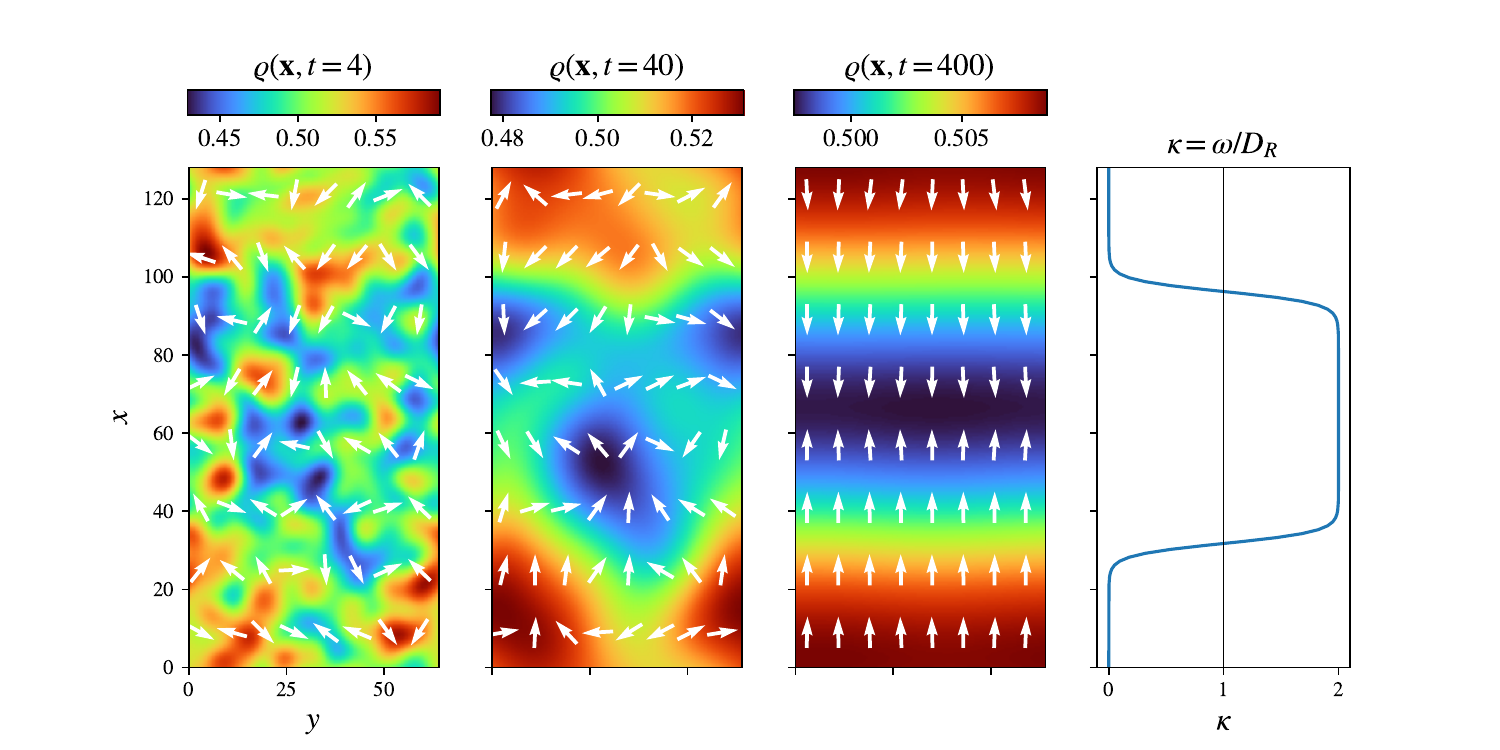}
    \caption{\textcolor{black}{We numerically solve the AMB+ dynamics of 
    Eq.~\eref{full_AMB_plus} 
    with the coefficients given by Eq.~\eref{AMB_plus_coeficients} for different 
    times on a periodic two-dimensional lattice \cite{zwicker2020pypde}. 
    We chose the system parameters for the passive and active diffusivity as 
    $D_T/D_0=1,\ \textcolor{black}{D_A^0}/D_0 = 0.05$
    rescaled by some reference diffusivity $D_0$, and the volume fraction as 
    $\phi = 0.1$. 
    With this choice of parameters we stay within the regime of validity of our 
    derived AMB+ dynamics, i.e., 
    $\mathrm{Pe} = \sqrt{2\textcolor{black}{D_A^0}/ D_T} \ll 1$ and $\phi\ll 1$. 
    Starting 
    from a random initial distribution, the density quickly relaxes to a 
    homogeneous steady state $\varrho(\mathbf{x},t) = \varrho_0 = 0.5$,
    where the white arrows indicate the direction of the normalised diffusive 
    flux 
    (note the different magnitudes of the density variation at different times).
    Different regions of active chirality $\kappa = \omega/D_R$ only affect the 
    density relaxation at 
    later times ($t = 40, 400$), as only higher-order density gradient terms of the 
    AMB+ are affected by the 
    sign-change of the coefficients. In the plots, time is rescaled by 
    the natural time scale $\tau = 1/D_R$ and space is measured in units of 
    $\sqrt{D_0\tau}$.}}
    \label{fig:AMB_plu_solution}
\end{figure}

\section{Conclusion}
\label{section_conclusion}

We here extended a geometric approach \textcolor{black}{\cite{bruna2022phase, 
bruna2012diffusion,bruna2012excluded,kalz2022collisions,kalz2022diffusion}} 
to deal with particle-particle interactions by restricting the domain of 
definition of \textcolor{black}{their diffusing centrers}. Forbidden overlaps of 
the particles correspond to forbidden areas in the domain, creating a 
configuration space with inner moving boundaries. Based on that we derived 
an effective time-evolution equation for the full one-body PDF. We proceeded 
by projecting the full one-body PDF onto its angular modes, which resulted 
in a coupled hierarchy of hydrodynamic modes. By scrutinising the underlying 
assumptions we turned the closure scheme into a perturbation problem and 
found effective time-evolution equations (field-theories) for the mean 
particle density. 

One major result of our work is the following. We observed that this procedure 
provides us with
first-principles access to the otherwise phenomenological parameters of 
field-theoretical descriptions of active matter. We find that, 
\textcolor{black}{beyond an effective diffusive description,} the 
microscopically best justified theory of a continuous model is the AMB+ 
\textcolor{black}{\cite{tjhung2018cluster, nardini2017entropy}}.
This was also reported in another recent first-principles
derivation of the AMB+ for non-chiral ABPs \cite{te2023derive}.
From a technical point of view, our work unravels the 
theoretical necessities of the regimes of validity to obtain such a 
coarse-grained model for the description of active matter. Specifically, 
we find that when neglecting the time evolution of higher modes such as 
polarisation or nematic order, in the so-called adiabatic limit, the AMB+ model 
is microscopically justified only in the limit of low P{\'e}clet numbers, i.e., 
when thermal diffusion dominates active motion. Whether the 
\textcolor{black}{microscopic justification for the} AMB+ model can be extended 
to regimes of higher activity is a subject of future research. 

The main prediction of this work is that active chirality has a 
non-trivial influence on the dynamics of the mean particle PDF 
$\varrho(\mathbf{x},t)$.
Even though in the simplest version of the ACP model chirality is not altered 
by particle interactions \textcolor{black}{(see again the Langevin description 
in Eq.~\eref{langevin_description})}
and hence its effect on the full one-body level is rather superficial, 
it becomes most prominent when integrating out the angular dependence of the 
full one-body PDF $p(\mathbf{x},\theta, t)$. We find that 
\textcolor{black}{an odd tensor \cite{kalz2022collisions,kalz2024oscillatory} 
$\boldsymbol{\Gamma}^{-1} \sim (\mathbf{1} - \kappa\boldsymbol{\varepsilon})$ 
emerges,
where chirality defines the off-diagonal elements $\kappa = \omega/D_R$. 
Powers of that tensor, and hence chirality, eventually,} can change 
the sign of all activity-induced coefficients of the AMB+, $\lambda$, and $\xi$,
as well as the Cahn-Hilliard coefficient 
$k\textcolor{black}{[}\varrho\textcolor{black}{]}$. 
\textcolor{black}{Restricted to the effective diffusion dynamics 
similar to Eq.~\eref{model_B}
the alternative approach of Ref. \cite{bickmann2022analytical} did not report 
this phenomenon, as their method to incorporate interactions assumes 
translational and 
rotational invariance and therefore can only treat chirality pertubatively}. 
\textcolor{black}{Whether the sign-change of the coefficients has implications 
for the phase-transition dynamics, however, cannot be addressed within our model due 
to its restricted validity to regimes away from phase transitions for particles 
with repulsive interactions.  
In the regime of validity of our AMB+ model, numerical solutions of 
Eq.~\eref{full_AMB_plus} with the parameters of 
Eqs.~\eref{AMB_plus_coeficients} only admit a
homogeneous phase, see Fig.~\ref{fig:AMB_plu_solution}.}

The systematic analysis presented in this work allows us to further include the 
effect of nematic order on the time-evolution of the mean particle PDF, but it 
would amount to a field-theoretical description at an even higher order than the 
AMB+ (which is already at fourth order in density gradients). 
\textcolor{black}{The AMB+ was recently shown to be deducible from such a 
(stable) 
higher-order model \cite{thomsen2021periodic}, where it was shown that the AMB+ 
itself is unstable for sufficiently high order parameters. We suspect a similar 
behaviour if one performed the corresponding closure scheme on the level 
of the nematic order in our theory, which we leave for future work.}

Inspired by a rich behaviour of complex macroscopic phenomena in active matter, 
\textcolor{black}{an additional} step would be to \textcolor{black}{directly} 
couple the chirality to the spatial interaction of the particles 
\cite{fruchart2021non, frohoff2021localized}. Motivated from an 
orientation-dependent potential of the form $\hat{\mathbf{e}}(\theta_i) 
\cdot \hat{\mathbf{e}}(\theta_j)$
\textcolor{black}{for the propulsion vectors of} particles $i$ and $j$ 
\cite{peruani2008mean},
we could consider additional aligning interactions together with active 
chirality from an analytic perspective 
\textcolor{black}{on the restricted domain resembling steric interactions}.
A recent work considered the numerical effects of such alignment for phenomena 
like motility-induced phase separation and the flocking of active chiral particles 
\cite{kreienkamp2022clustering}.

\ack

E. K. thanks Pietro Luigi Muzzeddu and Hidde Vuijk for 
valuable discussions on the topic. The authors further acknowledge support by 
the Deutsche Forschungsgemeinschaft (E. K., A. S. and R. M. trough DFG grants 
No. SPP 2332 - 492009952, SH 1275/5-1 and ME 1535/16-1)

\appendix

\section{Evaluation of integral \eref{seperated_integral_appendix}}
\label{appendix_integral_evaluation}

In the main text, in Eq.~\eref{seperated_integral_appendix}  we are left with 
evaluating a boundary integral, in which integration and differentiation are 
with respect to different particle labels. Hence a use of the Gaussian 
divergence theorem is not possible straightforwardly. Instead, we use an 
extended version of the Reynolds transport theorem, which usually allows for 
the time-differentiation of an integral quantity, where the integration 
volume $V$ itself is time-dependent $V = V(t)$. For an arbitrary space- and 
time-dependent function $f = f(\mathbf{x}, t)$, the theorem in this context 
reads 
\begin{equation}
    \frac{\partial}{\partial t} \int_{V(t)}\mathrm{d}\mathbf{x}\ f = 
    \int_{V(t)}\mathrm{d}\mathbf{x}\ \frac{\partial f}{\partial t} + 
    \int_{\partial V(t)}\mathrm{d}\mathrm{S}_\mathbf{x}\  \left(\mathbf{n} 
    \cdot \mathbf{v}_{\partial V(t)}\right)\ f,
\end{equation}
where $\mathbf{v}_{\partial{V(t)}} = \mathbf{v}_{\partial{V(t)}}(\mathbf{x},t)$ 
is the velocity of an element of the moving boundary $\partial V(t)$ and 
$\mathrm{d}\mathrm{S}_\mathbf{x}\ \mathbf{n}$ is the outward area element of 
the boundary at time $t$. 

This theorem can be extended to cases, in which the integration volume $V$ 
is space-dependent $(V = V(\mathbf{x})$, but of a constant shape) and we are 
interested in taking the divergence of the integral with respect to that 
coordinate $\mathbf{x}$. For a vector-valued function $\mathbf{f} = 
\mathbf{f}(\mathbf{x}, \mathbf{y}, t)$, the extended transport theorem reads 
\begin{equation}
\label{Reynolds_divergence_theorem}
    \nabla_\mathbf{x}\cdot \int_{V(\mathbf{x})}\mathrm{d}\mathbf{y}\ 
    \mathbf{f} = \int_{V(\mathbf{x})}\mathrm{d}\mathbf{y}\ \nabla_\mathbf{x}
    \cdot\mathbf{f} + \int_{\partial V(\mathbf{x})}
    \mathrm{d}\mathrm{S}_\mathbf{y}\  \mathbf{n}_\mathbf{y} \cdot \mathbf{f}.
\end{equation}
The proof of this relation can be found in Ref. \cite{kalz2022diffusion}. We 
will apply this theorem to the reduced configuration space $\Lambda(\chi_1) = 
\Omega\setminus \mathrm{B}_{\varepsilon_d}(\mathbf{x}_1) \times [0, 2\pi)$, 
where the space-dependence is on $\mathrm{B}_{\varepsilon_d}(\mathbf{x}_1)$, 
the disk of radius $\varepsilon_d$ centred at $\mathbf{x}_1$. Therefore, 
only $\partial\mathrm{B}_{\varepsilon_d}(\mathbf{x}_1)$ contributes a moving 
boundary, and hence a boundary integral in 
Eq.~\eref{Reynolds_divergence_theorem}. We now apply the extended Reynolds 
transport theorem to evaluate Eq.~\eref{seperated_integral_appendix}. Note 
here that $P_2(t) = P_2(\chi_1, \chi_2, t)$ for a shortness of notation. We 
find the result
\begin{eqnarray}
    \int_{\Lambda(\chi_1)}\mathrm{d}\chi_2\ \nabla_{\chi_1} \cdot 
    \bigg[\mathbf{D} \nabla_{\chi_1} - \mathbf{f}(\theta_1) \bigg] P_2(t) 
    \nonumber \\
    = \nabla_{\chi_1} \cdot \int_{\Lambda(\chi_1)}\mathrm{d}\chi_2\ 
    \bigg[\mathbf{D} \nabla_{\chi_1} - \mathbf{f}(\theta_1) \bigg] P_2(t) 
    \nonumber \\
    \label{to_be_combined_boundary_integral}
    \quad - \int_0^{2\pi}\mathrm{d}\theta_2\int_{
        \partial\mathrm{B}_{\varepsilon_d}(\mathbf{x}_1)} \mathrm{d}
        \mathrm{S}_{2}\ \mathbf{n}_{2}\cdot \bigg[D_T \nabla_{1} - 
        v\hat{\mathbf{e}}(\theta_1) \bigg] P_2(t).
\end{eqnarray}
The moving boundary only arises in the spatial part of $\chi_1 =
(\mathbf{x}_1, \theta_1)$, and thus we are only left with the spatial part 
of the diffusion matrix $\mathbf{D}=\mathrm{diag}(D_T, D_T, D_R)$ and drift 
term $\mathbf{f}(\theta_1) = (v\hat{\mathbf{e}}(\theta_1), \omega)^\mathrm{T}$. 
We again apply the extended theorem for the first term in the evaluated integral
\begin{eqnarray}
\label{evealute_boundary_term}
 \nabla_{\chi_1} \cdot \int_{\Lambda(\chi_1)}\mathrm{d}\chi_2\ \bigg[\mathbf{D} 
 \nabla_{\chi_1} P_2(t)\bigg] = \nabla_{\chi_1} \cdot \mathbf{D}\nabla_{\chi_1} 
 \int_{\Lambda(\chi_1)}\mathrm{d}\chi_2\ P_2(t) \nonumber\\
 \quad - \nabla_{1} \cdot \int_0^{2\pi}\mathrm{d}\theta_2\int_{\partial
 \mathrm{B}_{\varepsilon_d}(\mathbf{x}_1)} \mathrm{d}\mathrm{S}_{2}\ 
 \mathbf{n}_{2}\ \bigg[D_T\ P_2(t)\bigg].
\end{eqnarray}

We are left with evaluating the originating boundary integral combined with 
the boundary integral of Eq.~\eref{to_be_combined_boundary_integral}. Therefore 
we jointly use the divergence theorem and the transport theorem, paying respect 
to the specific integration volumes in the following steps. First we apply the 
divergence theorem on the volume $\Omega \setminus \mathrm{B}_{\varepsilon_d}
(\mathbf{x}_1)$ ``backwards'',
\begin{eqnarray}
\label{rein_raus_1}
    - \nabla_{1} \cdot \int_0^{2\pi}\mathrm{d}\theta_2\int_{
        \partial\mathrm{B}_{\varepsilon_d}(\mathbf{x}_1)} \mathrm{d}
        \mathrm{S}_{2}\ \mathbf{n}_{2}\bigg[D_T\ P_2(t)\bigg] = -D_T\nabla_{1} 
        \cdot \int_{\Lambda(\chi_1)} \mathrm{d}\chi_2\ \nabla_2 P_2(t) 
        \nonumber\\
    \quad  + D_T\nabla_{1} \cdot \int_0^{2\pi}\mathrm{d}\theta_2
    \int_{\partial \Omega}\ \mathrm{d}\mathrm{S}_{2}\ \mathbf{n}_{2} P_2(t).
\end{eqnarray}
Using the generalised transport theorem, the first integral on the right-hand 
side of Eq.~\eref{rein_raus_1} can be rewritten as
\begin{eqnarray}
\label{rein_raus_2}
-D_T\nabla_{1} \cdot \int_{\Lambda(\chi_1)} \mathrm{d}\chi_2\ \nabla_2 P_2(t) 
= -D_T \int_{\Lambda(\chi_1)} \mathrm{d}\chi_2\ \nabla_1\cdot\nabla_2 P_2(t) 
\nonumber \\
 \quad -D_T \int_0^{2\pi}\mathrm{d}\theta_2\int_{\partial
 \mathrm{B}_{\varepsilon_d}(\mathbf{x}_1)} \mathrm{d}\mathrm{S}_{2}\ 
 \mathbf{n}_{2}\cdot \nabla_2 P_2(t).
\end{eqnarray}
Note again here that only $\mathrm{B}_{\varepsilon_d}(\mathbf{x}_1)$ is 
space-dependent and thus contributes a boundary term in the generalised 
transport theorem. Since integration and differentiation are with respect to 
the same particle label in the first integral on the right-hand side of 
Eq.~\eref{rein_raus_2}, we can apply the divergence theorem ``forwards'' 
\begin{eqnarray}
\label{rein_raus_3}
-D_T \int_{\Lambda(\chi_1)} \mathrm{d}\chi_2\ \nabla_1\cdot\nabla_2 P_2(t) 
= -D_T \int_0^{2\pi}\mathrm{d}\theta_2 \int_{\partial\Omega} 
\mathrm{d}\mathrm{S}_2\ \mathbf{n}_2 \cdot \nabla_1 P_2(t)\nonumber\\ 
\quad -D_T \int_0^{2\pi}\mathrm{d}\theta_2 \int_{\partial 
\mathrm{B}_{\varepsilon_d}(\mathbf{x}_1)} \mathrm{d}\mathrm{S}_2\ 
\mathbf{n}_2 \cdot \nabla_1 P_2(t).
\end{eqnarray}
Note here that when applying the divergence theorem the box-boundary 
$\partial\Omega$ contributes a surface integral. As the box-boundary does not 
explicitly depend on $\mathbf{x}_1$, the partial differential operator 
$\nabla_1$ can be moved outside the integral (without creating another 
boundary term) and we observe that it cancels with the second integral on 
the right-hand side of Eq.~\eref{rein_raus_1}. We are thus left with the 
two integrals on the inner boundary $\partial \mathrm{B}_{\varepsilon_d}
(\mathbf{x}_1)$ and find that
\begin{eqnarray}
- \nabla_{1} \cdot \int_0^{2\pi}\mathrm{d}\theta_2\int_{
    \partial\mathrm{B}_{\varepsilon_d}(\mathbf{x}_1)} \mathrm{d}
    \mathrm{S}_{2}\ \mathbf{n}_{2}\bigg[D_T\ P_2(t)\bigg] \nonumber\\
= -D_T \int_0^{2\pi}\mathrm{d}\theta_2\int_{\partial 
\mathrm{B}_{\varepsilon_d}(\mathbf{x}_1)} \mathrm{d}\mathrm{S}_2\ \mathbf{n}_2 
\cdot \left(\nabla_1 + \nabla_2\right) P_2(t).
\end{eqnarray}

The rewriting of the boundary term of Eq.~\eref{evealute_boundary_term} finally 
enables us to combine the result with the term in 
Eq.~\eref{to_be_combined_boundary_integral}. Using the definition of the full 
one-body PDF $p(\chi_1, t) = \int_{\Lambda(\chi_1)}\mathrm{d}\chi_2\ 
P_2(\chi_1, \chi_2, t)$, Eq.~\eref{to_be_combined_boundary_integral} 
thus becomes
\begin{eqnarray}
    \int_{\Lambda(\chi_1)}\mathrm{d}\chi_2\ \nabla_{\chi_1} \cdot 
    \bigg[\mathbf{D} \nabla_{\chi_1} - \mathbf{f}(\theta_1) \bigg] P_2(t) 
    = \nabla_{\chi_1} \cdot \bigg[\mathbf{D} \nabla_{\chi_1} - 
    \mathbf{f}(\theta_1) \bigg] p(\chi_1, t) \nonumber \\
    \quad - \int_{\partial \mathrm{B}_{\varepsilon_d}(\mathbf{x}_1)} 
    \mathrm{d}\mathrm{S}_{2}\ \mathbf{n}_{2}\ \bigg[D_T \left(2\nabla_1 - 
    \nabla_2\right) + v\hat{\mathbf{e}}(\theta_1) \bigg] P_2(t),
\end{eqnarray}
which constitutes the result of Eq.~\eref{seperated_integral_appendix} 
in the main text.

\section*{References}
%\bibliographystyle{iopart-num}
%\bibliography{../../../parental_bibtex/mybib.bib}

\providecommand{\newblock}{}

\end{document}